# GPT Models in Construction Industry: Opportunities, Limitations, and a Use Case Validation

*Abdullahi Saka[1], Ridwan Taiwo[2], Nurudeen Saka[1], Babatunde Salami[3], Saheed Ajayi*[1], Kabiru Akande[4], and Hadi Kazemi[1]*

[1] School of Built Environment, Engineering and Computing, Leeds Beckett University, UK
[2] Department of Building and Real Estate, Hong Kong Polytechnic University, Hong Kong
[3] School of Computing, Engineering and Digital Technologies, Teesside University, UK
[4] OVO Energy, UK

* **Corresponding Author** (Professor in Digital Construction and Project Management; Construction Informatics and Digital Enterprise Laboratory (CIDEL))

**Graphical Abstract**

# GPT models in Construction Industry: Opportunities, Limitations, and a Use Case Validation.

## ① Introduction
GPT models are **Large Language Models (LLMs)** trained on large datasets and can perform well on diverse tasks in **Natural Language Processing (NLP)**. The review examines potential opportunities, emerging challenges and a Use Case validation of **GPT models** in the **construction industry**.

## ② Objectives
1. To identify opportunities for the application of **GPT models**
2. To evaluate the limitations to application of **GPT models** in the AEC industry
3. To validate a use case for **GPT models** in the AEC industry

## ③ Methodology
**Qualitative research approach** is adopted in this research to achieve the aim of the study. As such, three sequential steps are leveraged

## ④ Applications to Construction

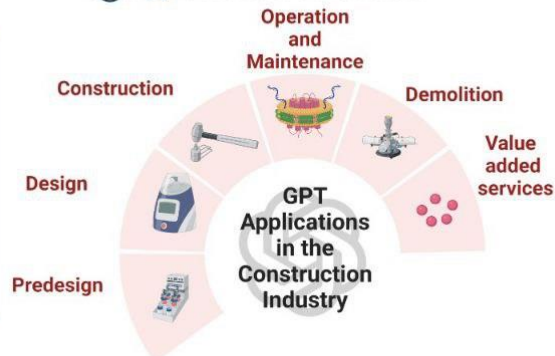

Operation and Maintenance, Demolition, Value added services, GPT Applications in the Construction Industry, Predesign, Design, Construction

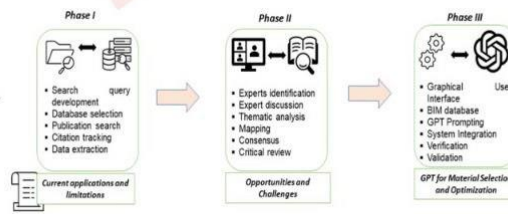

Phase I: Search query development, Database selection, Publication search, Citation tracking, Data extraction — *Current applications and limitations*
Phase II: Experts identification, Expert discussion, Thematic analysis, Mapping, Consensus, Critical review — *Opportunities and Challenges*
Phase III: Graphical User Interface, BIM database, GPT Prompting, System Integration, Verification, Validation — *GPT for Material Selection and Optimization*

## ⑤ Challenges
- Hallucinations
- Data and Interoperability
- Domain-specific knowledge and Regulatory compliance
- Confidentiality and Intellectual property
- And more...

## ⑥ Limitations
- **GPT models** could enhance productivity and efficiency in the construction industry
- Emerging challenges are inherent in the models and due to nature of the industry
- Research should overcome these barriers and leverage **GPT** in Construction

## ⑦ Case Study/Validations
- Material selection and Optimization Problem
- Data Retrieval Module
- NLP Prompt Processing Module
- User Interface and Integration Module

**Abstract**

Large Language Models (LLMs) trained on large data sets came into prominence in 2018 after Google introduced BERT. Subsequently, different LLMs such as GPT models from OpenAI have been released. These models perform well on diverse tasks and have been gaining widespread applications in fields such as business and education. However, little is known about the opportunities and challenges of using LLMs in the construction industry. Thus, this study aims to assess GPT models in the construction industry. A critical review, expert discussion and case study validation are employed to achieve the study's objectives. The findings revealed opportunities for GPT models throughout the project lifecycle. The challenges of leveraging GPT models are highlighted and a use case prototype is developed for materials selection and optimization. The findings of the study would be of benefit to researchers, practitioners and stakeholders, as it presents research vistas for LLMs in the construction industry.

## 1.0 Introduction

The architecture, engineering, and construction (AEC) industry is known for its slow adoption of innovation, when compared to other industries, due to the culture of the industry and the nature of its products (Gambatese & Hallowell, 2011). The industry is information-intensive and relies on myriad and diverse information from different stakeholders for successful project delivery (Chen & Kamara, 2005). However, there is a lack of information integration, reuse, and efficient management, all of which have a tremendous effect on stakeholders' collaboration and productivity of the industry. Past reports in the construction industry have emphasised the need for improvement in the modus Operandi of the industry to improve productivity and achieve value for money (Egan, 1998). Albeit the industry currently contributes about 13% to the global GDP, productivity growth has only been increasing at 1% per year over the last two decades (Ribeirinho *et al.*, 2020). Also, the industry is facing a myriad of challenges such as delays, health & safety, cost overrun, shortage of skilled personnel, and stringent requirements by governments. With the advancement in information technologies and digital tools, the AEC industry has been embracing its usage to improve its performance in a bid towards the fourth industrial revolution (industry 4.0). Consequently, there has been an increase in the usage of building information modelling (BIM), application of big data analytics, offsite construction, automation, and artificial intelligence (AI).

Artificial intelligence deals with the ability of machines to perform tasks that typically require human intelligence, such as learning, reasoning, perception and decision-making. AI systems process and analyse large datasets with the view of identifying patterns, relationships, drawing inferences, recommendations and taking action. Abioye *et al.* (2021) listed the subfield of AI to include machine learning, knowledge-based systems, computer vision, robotics, natural language processing, automated planning and scheduling and optimization. These diverse fields have been employed in the AEC industry to improve productivity and efficiency. As such, AI has been leveraged in cost prediction, delay prediction, building design energy prediction, workers' activity recognition, construction site safety, cash flow prediction, structural health monitoring, resource allocation and optimization, predictive maintenance, and decision support system, among others (Abioye *et al.*, 2021). Studies have shown significant improvement in productivity and efficiency with the use of these tools; however, bottlenecks have been reported. Challenges such as lack of skilled workers, cultural resistance to change, cost of implementation, unavailability of structured data, trust and ethics have been highlighted as the major hurdles towards the effective deployment of AI in the AEC industry (Akinosho et al., 2020).

With the advancement in AI, Conversational AI came of age in 2010 with the launch of Apple's Siri. Conversational AI deals with the application of NLP to enable computers to understand and interact with humans in a conversational way using natural language(Kulkarni *et al.*, 2019). This improved human-computer interactions and led to the development of chatbots, virtual assistants and other conversational interfaces that can assists users in automating tasks, information retrieval, customer services among others. Saka *et al.* (2023) reviewed the current applications of Conversational AI in the AEC industry and highlighted that the deployment of this emerging field is still limited. Few extant studies on the application of Conversational AI in the construction industry focused on information retrieval from BIM with limited functionalities. Majority of the developed Conversational AI agents in the AEC industry are based on the traditional approach to NLP, which requires time for processing the data, and users' interaction are often restricted as the agents are developed with the assumption of happy path users. Similarly, other few studies have leveraged on machine learning for the development of Conversational agents such as Bidirectional Encoder Representations from Transformers (BERT) and the use of commercial platforms such as IBM Watson, Amazon Alexa, Google Natural Language AI, and Microsoft Azure (Saka *et al.*, 2023). However, this approach often requires large data sets for training which are unavailable and expensive to gather in the AEC industry.

Furthermore, these machine learning approaches such as BERT are part of Large Language Models (LLMs) which came to limelight in 2018 after the introduction of transformer- a model architecture which rely on attention mechanism and differ from recurrent neural networks - by Vaswani *et al.* (2017). LLM are neural network with large parameters and trained using self-supervised learning and semi-supervised learning on large dataset. These LLMs have improved NLP and shifted the direction away from training with labelled data for defined objectives. Generative Pre-trained Transformer (GPT) models which are decoder blocks only from OpenAI and have gained significant attention and showed improved performance from GPT-2 (trained with 10 billion tokens) to GPT-3 (trained with 499 billion tokens) and recently GPT-4 released in 2023. The GPT models as a result of the large training dataset and large parameters have enabled few-shot (provide contexts and examples in the prompts), zero-shot (no example is provided in the prompt) learning capability (Wei *et al.*, 2022). As a result, it has been widely deployed in many applications. GPT models have beneficial applications in healthcare for triaging, analysing electronic health records, translation, medical education, medical and diagnostic (Li *et al.*, 2023). In bioinformatics, GPT model have been applied in sequence analysis, Genome analysis, Gene expression, proteomics, and in drug discovery (Zhang *et al.*, 2023). In education, GPT could transform autodidactic experience by providing personalized support, increased accessibility, flexible learning, real-time feedback and guidance (Firat, 2023). Similarly, GPT can be leveraged in business and commerce for chatbots, virtual assistants, customer service management, sentiment analysis, financial analysis and forecasting, fraud detection and supply chain management (Zong & Krishnamachari, 2022).

Despite GPT models overcoming some of the extant challenges of developing AI applications in the construction industry and providing opportunities to improve productivity, there are few studies on GPT models in the AEC industry. Also, there are no reviews on the opportunities of these emerging LLMs in the literature. Consequently, this current study aims to critically review GPT models in the AEC industry with the following objectives:

   a) To identify opportunities for the application of GPT models
   b) To evaluate the limitations to the application of GPT models in the AEC industry
   c) To validate a use case for GPT models in the AEC industry

Achieving these objectives would significantly contribute to the emerging body of knowledge on LLMs in the construction industry and provide research agenda for researchers. Also, this study highlights areas that would benefit from the application of GPT models and provides a case study validation for the use of GPT in material selection and optimization. Similarly, inherent challenges of the application of GPT are presented to enlighten stakeholders about the possible pitfalls that can be encountered in the deployment of GPT models and to prevent health, safety, and business problems. The rest of the paper is structured into seven sections (Figure 1) covering the literature review, methodology, findings, discussion, case validation and conclusion.

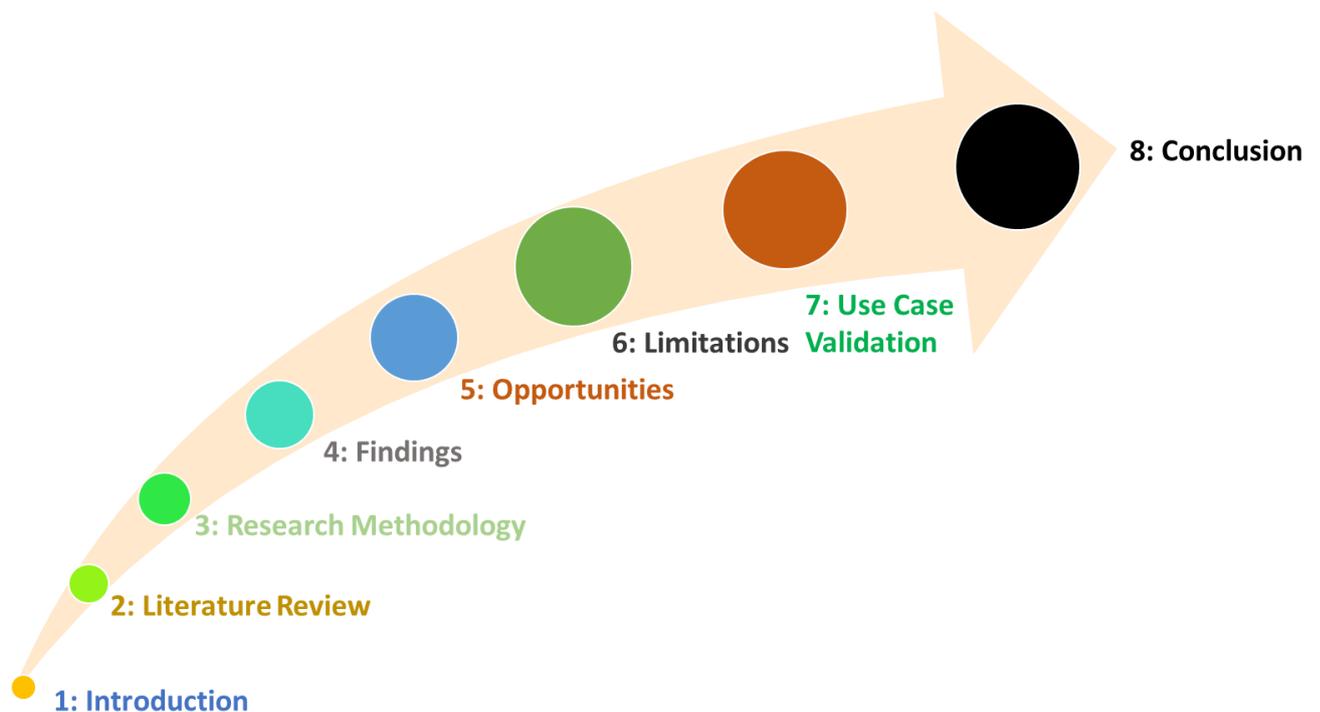

Figure 1: Structure of the Paper

## 2.0 Generative Pre-trained Transformer (GPT)

OpenAI's Generative Pre-trained (GPT) models have made significant contributions to the field of language generation. GPT models use transformer-based models that learn statistical patterns of natural language, enabling them to generate human-like language. The series started with GPT-1 in June 2018 and has since evolved to GPT-2, GPT-3, and GPT-3.5 (OpenAI, 2019; Radford et al., 2019). The latest addition, GPT-4, was launched in March 2023 and demonstrates significant advancements in generating coherent and understandable text. GPT models are trained using vast amounts of unstructured text data, enabling them to generate language almost indistinguishable from human-generated text (OpenAI, 2023). Early NLP relied on rule-based systems that required explicit programming of grammar rules and syntax. However, these systems have limitations in programming complex languages and linguistic nuances, making them less adaptable to new domains or contexts and less scalable (Shaalan, 2010). The rise of data-driven approaches, such as GPT models, enabled machine learning algorithms to learn from large amounts of data and recognize complex patterns in natural language without explicit programming of rules. The GPT-3 API was introduced in June 2020 and made publicly available in November 2021 (Karhade, 2022; OpenAI, 2020). It brought significant advancements in NLP technology making GPT-3 widely accessible. In January 2022, InstructGPT, a version of GPT 3.5, which can handle more complicated instructions, was released. In 2022, speech recognition software Whisper and GPT-3.5 upgrade to text-davinci-003 were introduced in September and November, respectively (Karhade, 2022; OpenAI, 2023; Radford et al., 2022). GPT-4 has further advanced the NLP after launching in March 2023, opening up new possibilities for industry-specific applications. The GPT models have transformed the field of NLP, enabling previously unattainable levels of fluency and coherence in machine-generated text (OpenAI, 2023). **Table 1** provides a chronological summary of significant milestones in the development and release of GPT models.

**Table 1.** The progression of OpenAI's GPT models from GPT-1 to GPT-4.

| Date | Milestone | References |
|---|---|---|
| June 11th, 2018 | OpenAI introduced GPT-1, the first model in the GPT series. | (MUO, 2023) |
| Feb 14th 2019 | OpenAI announced the release of GPT-2. | (Radford et al., 2019) |
| May 28th 2020 | OpenAI published the initial GPT-3 preprint paper on arXiv. | (Brown et al., 2020) |
| June 11th 2020 | OpenAI launched a private beta for the GPT-3 API. | (VentureBeat, 2020) |
| Sep 22nd 2020 | OpenAI licensed GPT-3 to Microsoft. | (OpenAI, 2020) |
| Nov 18th 2021 | The GPT-3 API was opened to the public. | (VentureBeat, 2020) |
| Jan 27th 2022 | OpenAI released InstructGPT as text-davinci-002, later renamed GPT-3.5. | (OpenAI, 2023) |
| Jult 28th 2022 | OpenAI published a paper on exploring data-optimal models with FIM. | (Bavarian et al., 2022) |
| Sep 1st 2022 | OpenAI reduced the pricing of the GPT-3 model by 66% for davinci model. | (Decoder, 2022; OpenAI, 2022) |
| Sep 21st 2022 | OpenAI announced Whisper (speech recognition). | (Radford et al., 2022) |
| 28/Nov/2022 | OpenAI expanded GPT-3.5 to text-davinci-003 with improved language generation capabilities. | (Karhade, 2022) |
| Nov 30th 2022 | OpenAI announced ChatGPT. | (OpenAI, 2022) |
| Mar 14th 2023 | OpenAI released GPT-4, the latest and highly anticipated addition to the GPT series. | (OpenAI, 2023) |

One of the main advantages of GPT models is their capacity to produce language that is cohesive, fluent, and nearly indistinguishable from text produced by humans. These models have been effectively used in a variety of applications, including chatbots, content generation, and machine translation. They can produce answers to open-ended questions, making them an important tool for natural language communication. The layers of GPTs' transformer-based neural architecture employ attention techniques to concentrate on particular areas of the input text (Neelakantan et al., 2022; Vaswani et al., 2017). The model can pick up on statistical patterns in natural language attributable to its architecture without having to explicitly program it with syntax or grammar rules. The transformer network creates coherent and fluent output while the attention mechanism enables the model to focus on pertinent portions of the input text (Zhang et al., 2022). Transformer networks, feedforward neural networks, and attention processes make up the building blocks of GPTs (Hernández & Amigó, 2021). Massive volumes of text data are trained during the pre-training phase of GPTs, allowing the model to learn broad language patterns that may be honed for particular tasks (Kotei & Thirunavukarasu, 2023). Pre-training often involves unsupervised learning without labels or annotations. After pre-training, the model is adjusted for a variety of tasks to increase the quality and accuracy of the text that is produced for that activity, such as language modelling, text categorization, or question-answering.

GPT models may be fine-tuned to accomplish a range of tasks with great accuracy (Ouyang et al., 2022; Wei et al., 2023). GPTs are capable of a variety of tasks such as language production, sentiment analysis, text categorization, and question answering (Brown et al., 2020). Language generation is the process of creating coherent and fluent text, whereas sentiment analysis is the examination of text sentiment, such as whether it is positive or negative. Text categorization entails classifying text into several groups, such as news articles or product reviews. GPTs are useful tools for natural language interaction because question-answering creates replies to open-ended questions (Brown et al., 2020). GPTs have transformed the area of NLP, allowing hitherto impossible levels of fluency and coherence in a machine-generated text (Devlin et al., 2018; Radford et al., 2018). The likelihood of bias in the pre-training data, which may affect the accuracy and quality of the produced text, is one of the major difficulties. Additionally, the deployment and training of GPTs can be computationally difficult and expensive (Mantel, 2023). The field of GPTs is continuously changing as researchers explore new approaches to increase the precision and effectiveness of these models. This is due to the prognosis for future lines of study. For instance, the computational cost of developing and implementing GPTs may be reduced, making them more affordable for smaller businesses (Hussin et al., 2023; Paaß & Giesselbach, 2023). The problem of bias in pre-training data, which might affect the precision and standard of produced text, is another field of investigation (Paaß & Giesselbach, 2023).

## 3.0 Methodology

Qualitative research approach is adopted in this research to achieve the aim of the study. This involved leveraging three sequential steps, as depicted in Figure 2, including literature review, critical review, expert discussion, and a case study.

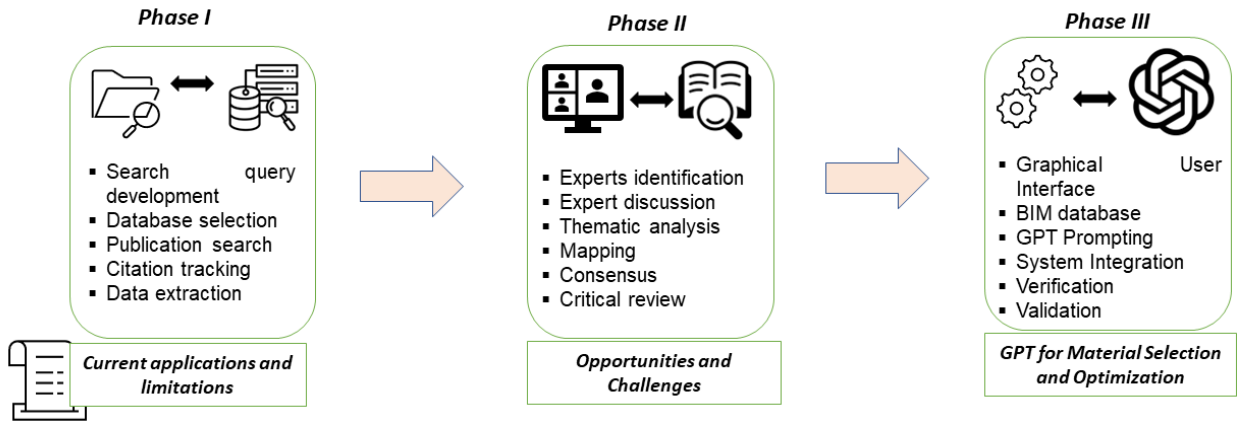

Figure 2: Research Approach

i) Initial Exploration (Phase I): A search query (shown in Table 2) is developed and used in conducting a detailed search in Scopus, ACM, Web of Science, Science Direct, and Google Scholar databases. These databases were selected because of their capacity, and relevance to the subject matter and have been well-adopted in similar studies (Saka *et al.*, 2023). The outputs from all the databases revealed that, although transformers have been gaining increased attention in the AEC industry, the application of GPT models is low. As such, only two publications were identified that specifically leveraged GPT models in the AEC industry. Furthermore, a subsequent search on arXiv database suggested that there are 3 related preprints – manuscripts that are yet to be formally peer-reviewed – available online. All the identified outputs were reviewed and citation tracking – getting previous studies from references - is used till saturation. These outputs are tabulated in an Excel file and considered to avoid publication bias. Lastly, the outputs were reviewed, and themes related to the objectives were identified and tabulated for usage in the subsequent research phases.

Table 2: Search query

| Search Category | Search Query |
|---|---|
| Construction Industry | "Construction industry" OR "architecture engineering and construction industry" OR "AEC industry" OR "AECO industry" |
| GPT | "Generative Pre-trained Transformers" OR "GPT" OR "GPT-1" OR "GPT-2" OR "GPT-3" OR "InstructGPT" OR "ChatGPT" OR "Transformer*" OR "GPT-4" |

ii) Expert Discussion and Critical Review (Phase II): Based on the outputs of Phase I, expert discussion and a detailed critical review were employed to complement the few research studies. Expert discussion is a method of obtaining in-depth insights on a specific research theme by facilitating a structured discussion with a group of panellists with diverse backgrounds and experiences. It is a strong approach when the research area is new or ambiguous and there is a need for the generation of innovative ideas, identifying problems, develop recommendations and solutions to complex problems. It could take various forms such as a Delphi survey or Focus Group Discussion and involves a moderator to ensure group dynamic and round participation from all the experts (Hsu & Sandford, 2007; Jenkins & Smith, 1994). As such, the employed expert discussion is a modified classical Delphi survey with panellists who are selected based on predefined criteria - domain expertise in Artificial Intelligence and the AEC industry with a minimum of 10 years' experience. Ten experts were identified and contacted for the research discussion and only 7 accepted the invites and participated in the research. This is considered acceptable as a minimum of 7 is considered sufficient for the Delphi survey and this current study leverages the expert discussion to complement critical review and case study validation (Hon *et al.*, 2011). Table 3 shows the demographic details of the experts that participated in the discussion. The experts were contacted virtually, and an average discussion of 25 minutes was undertaken on opportunities and limitations of GPT models in the AEC industry. Thematic analysis and mapping were conducted to identify specific opportunities mentioned by the experts. Opportunities and limitations were tabulated and pass across back to all the panellists for final review.

Following the expert discussion, a critical review was conducted on the identified opportunities and limitations of GPT models in the AEC industry. A critical review entailed a detailed analysis and critique of the work with the view of providing an objective assessment of the work, implications, and insights. It is employed in this study to evaluate and identify the opportunities and limitations of GPT models in the AEC industry based on the expert discussion and relevant extant studies.

Table 3: Experts' Demographic Details

| Designation | Professional Background | Experience | Sector and Expertise |
|---|---|---|---|
| A | Architect | 15 years | Research and Development with expertise in the deployment of Artificial Intelligence |
| B | Software Developer | 14 years | IT with expertise in developing solutions for AEC companies |
| C | Civil Engineer | 12 years | Research and Development with expertise in the deployment of Artificial Intelligence |
| D | Project Manager | 14 years | Research and Development with expertise in construction project analytics |
| E | Computer Engineer | 19 years | Research and Development with expertise in the deployment of Artificial Intelligence in the Construction Industry |
| F | Architect | 19 years | Research and Development with expertise in business intelligence |
| G | AI/ML Engineer | 10 years | IT with expertise in developing solutions for AEC companies |

iii) Use case (Phase III): One of the identified opportunities of GPT models in Phase II – material selection and optimization – is evaluated in this study. The system architecture is proposed for leveraging GPT for material selection and optimization which is subsequently verified and validated. Verification deals with 'building the product right' and validation deals with 'building the right product' (Boehm, 1984). The verification and validation process is done using checklists and case-testing approach (Saka *et al.*, 2022). The checklist technique involves the use of specialized lists based on experience to check significant issues that are critical for product development whilst case-testing technique entails prototyping the developed products. As such, the employed techniques in this study fulfil the basic criteria of verification and validation processes – completeness, consistency, feasibility and testability (Boehm, 1984).

## 4.0 Findings

This section presents the findings from the literature search on the current applications of GPT models in the AEC industry and expert discussion on opportunities and limitations.

### 4.1 Current applications

GPT models are still new in the construction industry unlike other large language models (LLM) such as BERT which has been gaining widespread applications in the AEC industry since 2020. Only 4 papers that have applied GPT models were retrieved and summarized in Table 4.

Table 4: Current Applications

| S/N | Application | Purpose | Access and GPT model (Specification) | Limitations |
|---|---|---|---|---|
| 1 | BIM-GPT (Zheng & Fischer, 2023) | Information retrieval from BIM using natural language | API access and GPT-3.5-turbo (Temperature = 0) | No quantitative evaluation Single-turn conversation |
| 2 | RoboGPT (You *et al.*, 2023) | Automate sequence planning of construction tasks in robot-based assembly | ChatGPT-4 API access (Not provided) | Blackbox High risk Unable to leverage visual information |
| 3 | (Prieto *et al.*, 2023) | Scheduling of construction tasks | ChatGPT-3.5 (Not provided) | Zero-shot learning No quantitative evaluation |
| 4 | (Amer *et al.*, 2021) | Integrating master schedules with look-ahead plans | GPT-2 (small version) | Large data set |

These extant studies have applied GPT models for information retrieval from BIM model and for scheduling and sequencing tasks. There are inherent limitations as a result of the GPT models employed in the applications and limitations as a result of the development approach employed in the studies. GPT models work best with well-structured and clean data, which is often unavailable in the AEC industry; as such, there is a need to pre-process the data before usage in GPT. Unstructured data are parsed into readable formats like Plain text files (.txt), Comma-separated values (.csv), and JavaScript Object Notation (.json). For instance, Zheng *et al.* (2023) employed BSON (Binary JSON) format by extracting building objects and properties from BIM model which is subsequently cleaned and stored in MongoDB. On the other hand, Amer *et al.* (2021), Prieto *et al.* (2023) and You *et al.* (2023) leveraged text for the application. Despite some of the limitations, the studies are important and contribute to the new area of GPT applications in the AEC industry and provide a basis for new studies to build on.

## 4.2 Expert Discussion

Table 5 summarizes the opportunities for GPT models in the AEC industry as identified through expert discussion, and they are categorized into different phases of project lifecycle.

Table 5: Categorization of Opportunities by Experts

| Phase | Opportunities | A | B | C | D | E | F | G |
|---|---|---|---|---|---|---|---|---|
| **Pre-Design** | Optimal design and construction techniques | ✓ | ✓ | ✓ |  | ✓ | ✓ | ✓ |
|  | Procurement |  | ✓ | ✓ | ✓ | ✓ |  |  |
|  | Project brief and client requirements | ✓ |  | ✓ |  |  |  | ✓ |
|  | Lessons from project | ✓ | ✓ | ✓ |  | ✓ |  | ✓ |
|  | Project execution planning | ✓ | ✓ |  |  |  |  | ✓ |
|  | Project management and planning |  |  |  | ✓ |  | ✓ | ✓ |
| **Design** | Generation of design concept |  | ✓ |  | ✓ |  |  |  |
|  | Regulatory compliance |  | ✓ |  | ✓ |  |  |  |
|  | Material selection and optimization | ✓ |  |  | ✓ | ✓ | ✓ | ✓ |
|  | Quantity take-off and costing | ✓ |  |  | ✓ | ✓ | ✓ | ✓ |
|  | Improving energy efficiency analysis | ✓ |  |  |  | ✓ | ✓ | ✓ |
|  | Design specification |  | ✓ |  |  |  | ✓ |  |
| **Construction** | Scheduling and logistics | ✓ | ✓ | ✓ | ✓ | ✓ | ✓ | ✓ |
|  | Regulatory compliance | ✓ | ✓ | ✓ | ✓ | ✓ | ✓ | ✓ |
|  | Risk identification, assessment, and management | ✓ | ✓ | ✓ | ✓ | ✓ | ✓ | ✓ |
|  | Progress monitoring and report | ✓ | ✓ |  |  | ✓ |  | ✓ |
|  | Site safety management | ✓ | ✓ | ✓ | ✓ | ✓ | ✓ | ✓ |
|  | Resource allocation and optimization | ✓ | ✓ | ✓ | ✓ | ✓ | ✓ | ✓ |
|  | Change order management |  | ✓ |  | ✓ |  |  | ✓ |
|  | Quality control and assurance |  |  |  |  | ✓ |  | ✓ |
|  | Documentation | ✓ | ✓ | ✓ | ✓ | ✓ | ✓ | ✓ |
|  | Dispute resolution |  | ✓ |  |  |  |  |  |

| | | A | B | C | D | E | F | G |
|---|---|---|---|---|---|---|---|---|
| | Budgeting and cost planning | | | ✓ | | | | |
| **Operation and Maintenance** | Predictive maintenance | | | | ✓ | | ✓ | |
| | Energy management and optimization | ✓ | | | ✓ | | | |
| | Incident and resolution | ✓ | ✓ | ✓ | ✓ | ✓ | ✓ | ✓ |
| | Lifecycle management of asset | ✓ | ✓ | ✓ | ✓ | ✓ | ✓ | ✓ |
| | Occupant communication and support | ✓ | ✓ | ✓ | ✓ | ✓ | ✓ | ✓ |
| | Regulatory compliance management | ✓ | ✓ | ✓ | ✓ | ✓ | ✓ | ✓ |
| | Space/facility management | ✓ | ✓ | ✓ | ✓ | ✓ | ✓ | ✓ |
| | Performance monitoring | ✓ | | | ✓ | | | |
| | Sustainability | | | | ✓ | | ✓ | |
| | Waste management and recycling | ✓ | ✓ | | | | | |
| **Demolition** | Demolition protocol | ✓ | ✓ | ✓ | ✓ | ✓ | ✓ | ✓ |
| | Waste management | | | | | | | |
| | Redevelopment plan | ✓ | ✓ | ✓ | ✓ | ✓ | ✓ | ✓ |
| | Regulatory compliance and permit | ✓ | ✓ | ✓ | ✓ | ✓ | ✓ | ✓ |
| | Costing | | | | ✓ | | | |
| | Environmental impact analysis | | ✓ | | ✓ | ✓ | | |
| | Material recovery | | | | | ✓ | ✓ | |
| | Risk assessment | ✓ | ✓ | | | | | |
| **Value added** | Knowledge management and training | ✓ | ✓ | ✓ | ✓ | ✓ | ✓ | ✓ |
| | Customer services | ✓ | ✓ | ✓ | ✓ | ✓ | ✓ | ✓ |
| | Stakeholder communication | | ✓ | ✓ | | ✓ | | |
| | Business intelligence | ✓ | ✓ | ✓ | | ✓ | ✓ | ✓ |
| | Conversational AI/Chatbot | ✓ | | ✓ | ✓ | ✓ | | ✓ |

Table 6 presents the challenges and limitations of deploying GPT models in the AEC industry. Some of these limitations are inherent limitations of GPT models, whilst others are industry-specific challenges facing the deployment of GPT models in the AEC industry.

Table 6: Limitations of GPT in the Construction Industry

| Limitations | A | B | C | D | E | F | G |
|---|---|---|---|---|---|---|---|
| Hallucination | ✓ | ✓ | | | ✓ | | ✓ |
| Data | ✓ | ✓ | ✓ | ✓ | ✓ | | |
| Expertise knowledge | ✓ | ✓ | ✓ | ✓ | ✓ | ✓ | |
| Intellectual property and confidentiality | | ✓ | | ✓ | | | |
| Safety | ✓ | ✓ | ✓ | ✓ | ✓ | ✓ | ✓ |
| Trust and acceptance | ✓ | ✓ | ✓ | ✓ | ✓ | ✓ | ✓ |
| Accountability and liability | | | | ✓ | | | |
| Ethics | ✓ | ✓ | ✓ | ✓ | ✓ | ✓ | ✓ |
| Skills and training | | | ✓ | ✓ | ✓ | ✓ | ✓ |
| Interoperability and integration | | | ✓ | | | | |
| Capital/cost | ✓ | ✓ | ✓ | ✓ | ✓ | ✓ | ✓ |
| Infrastructure requirements | | | ✓ | | | ✓ | ✓ |
| Scalability | ✓ | | ✓ | ✓ | | | |
| Performance optimization | | ✓ | | | ✓ | | ✓ |
| Cybersecurity | ✓ | ✓ | | ✓ | | ✓ | |
| Interdisciplinary | | | ✓ | | | | |
| Cultural and social consideration | ✓ | | ✓ | | ✓ | ✓ | |
| Latency issue | | | | | | | |

The following Sections 5 and 6 provide a detailed discussion of the opportunities and limitations identified in the previous section. Opportunities and limitations of similar themes are synthesised and discussed. These opportunities require different levels of development and interaction with the GPT models ranging from zero-shot learning to finetuning and integration with different knowledge sources. Similarly, some of the limitations are inherent in the GPT models whilst others can be attributed to the construction industry context.

## 5.0 Opportunities

The opportunities are categorized into different phases of the construction project lifecycle – predesign, design, construction, operation & maintenance, demolition phase and value-added services. The following subsections present these opportunities and how GPT models can be leveraged.

### 5.1 Pre-Design Phase

Pre-design construction operations include site analysis, programming, construction cost analysis, and value engineering to provide the framework for project success, budgeting, and optimization. This section explores the existing state of pre-design in construction projects and the opportunities for leveraging GPT models as shown in Figure 3.

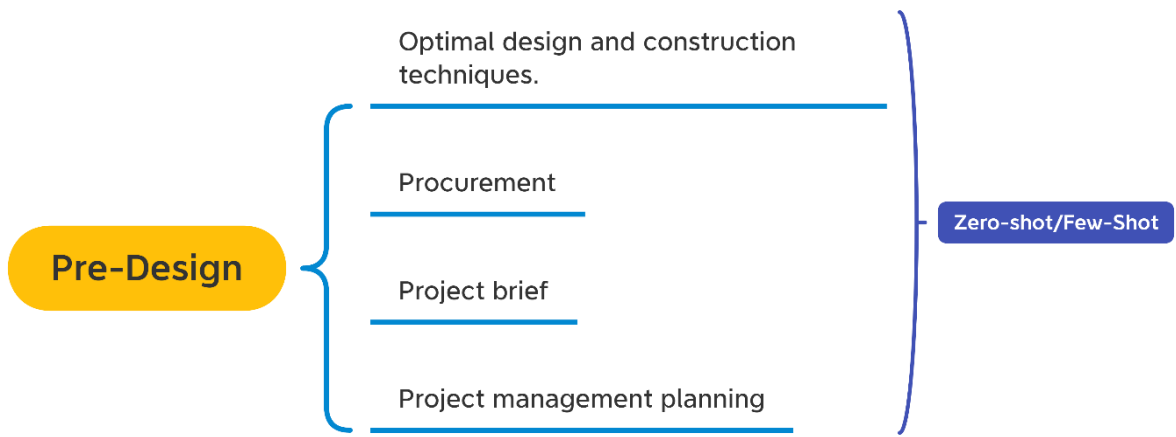

Figure 3: Opportunities for GPT models in Pre-design Phase

#### 5.1.1 Expert advice on optimal design and construction techniques.

Construction's pre-design phase is a crucial time when stakeholders work together to set project goals and deadlines. Pre-design's purpose is to identify critical aspects and restrictions that will guide subsequent design and development processes. Currently, the predesign landscape is defined by a mix of traditional approaches and specialized technologies, each with its own set of benefits and drawbacks. Due to cognitive biases and information overload, the traditional reliance on expert knowledge and experience might result in less-than-ideal outcomes (Levy, 2010). Extant studies have documented the importance of predesign in reducing project risk, optimizing resource allocation, and enhancing overall project success (Guo & Zhang, 2022; Lu et al., 2020; Obi et al., 2021). This phase can further be enhanced and transformed by GPT models, an AI-driven language model, which provides expert guidance on the best design and building procedures, resulting in higher productivity, decreased costs, and improved sustainability (OpenAI, 2023). Users may access the model's knowledge using the user interface or API access, promoting cooperation between project participants and guaranteeing that design and construction choices are in line with project objectives (Smith et al., n.d.).

Incorporating GPT models into predesign processes can help to simplify decision-making, enhance communication among stakeholders, and speed up the discovery of design restrictions and possibilities. Furthermore, the adoption of these models can reduce human error and biases during the predesign process, ensuring that projects are better aligned with objectives and requirements. GPT models will improve pre-design's contextual knowledge and produce appropriate suggestions after training on project-specific data. It is necessary to handle issues including data privacy, concerns about security, and potential biases in AI-generated recommendations for improvements. As AI and machine learning continue to progress, new prospects for innovation, sustainability, and project success are anticipated (OpenAI, 2023; Radford et al., 2022; Smith et al., n.d.).

### 5.1.2 Procurement decision support

In the predesign phase, current procurement decision support procedures include techniques such as value engineering, life cycle costing, and multi-criteria decision-making (MCDM) (Galjanić et al., 2022; Tezel & Koskela, 2023). These techniques, however, are constrained by their dependence on expert knowledge, which is frequently subjective and sensitive to human mistakes (Ratnasabapathy & Rameezdeen, 2010). Furthermore, existing research has emphasized the need for more effective data processing and analysis to improve decision-making, particularly with the growing volume of project data collected during the predesign phase (Budayan et al., 2015;).

By overcoming the limitations of traditional methods, GPT models provide a breakthrough approach to procurement decision assistance during the predesign phase (Mcbride et al., 2021). GPT models, with their powerful natural language processing and machine learning capabilities, can evaluate data, delivering insightful suggestions that encourage accurate, data-driven decisions (OpenAI, 2023). GPT models also promote more efficient cooperation among stakeholders through real-time information and predictive analysis (Abioye et al., 2021; Momade et al., 2021). Prompts can be developed allowing procurement decision-makers to swiftly acquire important information, explore options, and make educated decisions.

### 5.1.3 Development of project brief and client requirements

A crucial part of the predesign stage of the construction process is the development of the project brief and client requirements, often known as Employer Information Requirements (EIR) in BIM projects (Catenda, 2020; Kim et al., 2022). Creating a thorough project brief and outlining the client's requirements in detail are essential stages that provide the groundwork for the whole construction process at the pre-design stage of the project. Multiple methods and constraints, such as the use of templates, checklists, and questionnaires, are part of the existing ways of collecting design requirements (Assaf et al., 2023).

Numerous approaches and restrictions that affect the creation of the project brief and employer information requirements (EIR) have been highlighted by recent studies. For instance, several studies have stressed the value of stakeholder engagement, which entails speaking with customers, end users, and other stakeholders to determine their requirements and preferences. Other studies have emphasized the importance of information management systems, which can enhance client and project team cooperation and communication (Akinradewo et al., 2023). There are still several aspects that require improvement, such as the requirement for the EIR to include sustainability and resilience concepts (Ashworth et al., 2016). There are several areas that necessitate further development in the construction industry. These include the need for a client-centric approach, increased stakeholder engagement, and more efficient data collection and management systems (Al-Reshaid et al., 2005). By automating some procedures and fostering cooperation and communication between project teams and clients, the emergence of GPT models has the potential to completely transform the development of project briefs and EIRs (Shaji George et al., 2023). By analyzing data and locating pertinent information using natural language processing techniques, GPT models may help create project briefs. Furthermore, by offering a platform for real-time collaboration and feedback, GPT models can facilitate communication between project teams and clients (Di Giuda et al., 2020). Project teams may produce more precise and thorough project briefs and EIRs that consider the needs and preferences of the customer by utilizing GPT (Parm AG, n.d.).

### 5.1.4 Project management planning

Project management planning is vital to the successful delivery of projects, as poor planning could lead to cost & time overrun and quality issues (Asiedu *et al.*, 2017). It entails inputs from project team and stakeholders and set out details of how the project is to be executed, monitored, and controlled. As such project management planning components are baselines for scope, schedule, cost, requirement management plan, change management plan, configuration management plan and process improvement plan (Globerson & Zwikael, 2002).

GPT models trained on corpus of database could be finetuned to support project management planning during predesign stage in scope definition, scheduling, resource allocation & estimation, risk management and decision support system. For instance, with prompting, GPT models can help understand the impact of diverse factors on projects and provide recommendations for effective management. Also, GPT models have been proved applicable in the identification of tasks and scheduling based on specific project requirements (Prieto *et al.*, 2023; You *et al.*, 2023). As such, GPT models could also be leveraged for analysing data and providing relevant insights to the project manager during the predesign stage. Similarly, with zero-shot learning, GPT can act as a repository for industry standards and lessons learnt from previous projects for stakeholders to make informed decisions.

## 5.2 Design Phase

This section presents the opportunities for leveraging GPT models during the design phase of construction projects as shown in Figure 4.

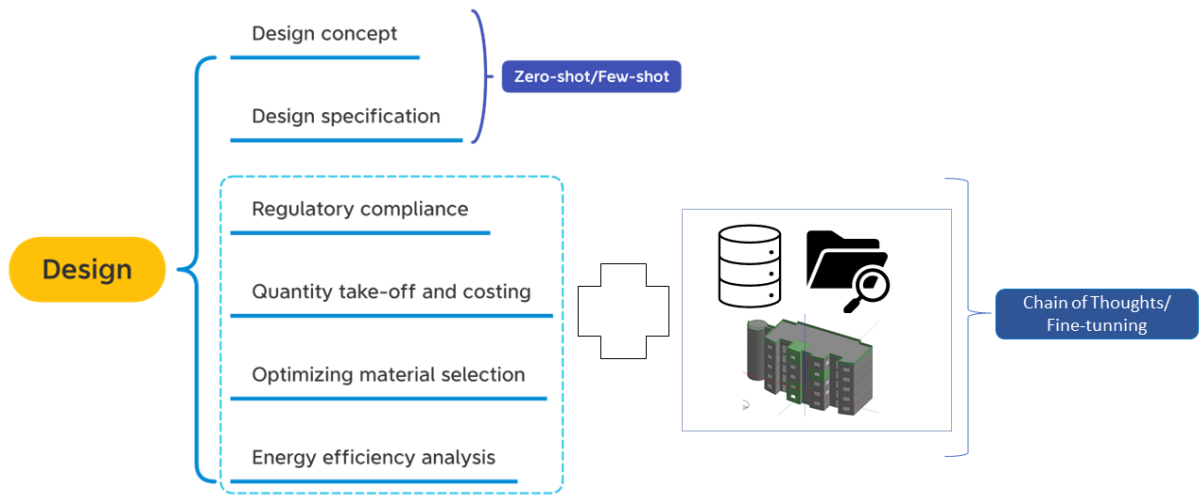

Figure 4: Opportunities for GPT models in Design Phase

### 5.2.1 Generation of Design Concept

One of the fundamental tasks during the design phase is to generate design concepts or alternatives that meet the specific project constraints. Traditionally, architects and engineers rely on their expertise and experience to develop these concepts, which can be a time-consuming and subjective process (As et al., 2018; Lewis & Séquin, 1998). However, GPT models offer a promising solution to enhance and expedite this process.

GPT models have demonstrated remarkable capabilities in generating coherent and contextually relevant text, making them well-suited for generating design concepts based on given project constraints (Zheng & Fischer, 2023). By training the model on a vast corpus of design-related texts, such as architectural plans, construction standards, and design guidelines, GPT models can learn the language and patterns specific to the construction industry.

### 5.2.2 Automated Regulatory Compliance

Complying with regulatory requirements is a critical aspect of the design phase, as failure to meet these standards can result in legal issues, delays, and safety hazards. The complexity and frequent updates of building codes and regulations make it challenging for design professionals to ensure compliance manually (Dimyadi et al., 2015). Nevertheless, GPT models can play a significant role in automating regulatory compliance checks, reducing errors, and streamlining the design process. This can be achieved by analysing architectural and structural designs and comparing them against relevant building codes. Conversely, the model can identify potential issues, allowing designers to rectify them before construction, thereby saving time and minimizing costly revisions.

### 5.2.3 Optimizing material selection.

The construction industry consumes a lot of resources and contributes significantly to the green gases emission. Material selection is one of the ways to reduce the environmental impacts of project during its lifecycle. Factors such as cost constraints, location of the component, design consideration and environmental requirements are decision-making factors for material selection (Florez & Castro-Lacouture, 2013). During the selection, the designer would need to consider all these factors coupled with other subjective and objective measures to reach a decision which may not be the optimal solution. As such, extant studies have proposed the use of different optimization approaches such as mixed integer optimization, fuzzy logic approach, and Grey relational analysis to solve the problem (Emovon & Oghenenyerovwho, 2020). GPT models can be leveraged for optimization of material selection with detailed consideration of different factors and parameters. This can be integrated with BIM model to optimize material selection during design and to provide design alternatives. It would involve evaluating the material of building components (from BIM) based on material properties and performance database (from GPT) to fulfil predefined criteria. This is further demonstrated in Section 7 for validation.

### 5.2.4 Quantity take-off and costing

Quantity take-off and costing form an integral part of successful project delivery, and the quantity surveyor is often saddled with this responsibility (Saka & Chan, 2019). GPT models can be leveraged for quantity take-off and costing of building projects by providing detailed information about the project – design, material and other specifications. This can be achieved by proving textual data for the model with necessary cost databases and estimation methods. Based on these, prompts can be developed for quantity take-off of the project with elemental breakdown and subsequent costing of the quantities. Also, GPT can be used to prepare bills of quantities and analysed past bills of quantities to draw insights and make predictions. However, due care should be taken to avoid relying on hallucinated outputs and the GPT should rather be used as a complementary tool for costing. Also, due to the current input format of GPT models, these models cannot take off building design automatically, except by providing textual data and specifications of the building.

### 5.2.5 Improving energy efficiency analysis.

There has been a surge in energy demand with rapid development over the decades in the construction industry (Oluleye *et al.*, 2022). The energy efficiency of buildings are is broadly based on the building envelope, which influences energy consumption and the rate at which energy is lost in the building (Abu Bakar *et al.*, 2015). GPT models can be provided with information on standards, regulations, passive design principles, building facades optimization and renewable energy systems for it to be leveraged for improving energy efficiency analysis. As such, GPT models can provide guidance on selection of simulation tools, interpretation of results, identify opportunities for improvement (such as optimizing building orientation, insulation, energy systems) and analysing cost-effectiveness of proposed solutions. Also, life cycle analysis can be computed with prompts developed for usage on the finetuned model and documentations can also be prepared by the GPT models. Whilst some of these are possible with zero-shot learning, to get best output from the GPT models, there is need to fine-tune the model and integrate it with other BIM and energy efficiency analysis tools.

### 5.2.6 Design specification

As one of the largest industries with in-built complexities, design specification plays a crucial role in the success of any project within the construction industry. Design specification, being a technical document, provides all necessary information for the design phase of a construction process. The existing process for developing a design specification document entails manual work, which can be time-consuming and laborious. However, with the introduction of the GPT models, new prospects for automating the process of developing design specification documents have surfaced. According to recent research, GPT models can generate logical and grammatically accurate language, which can be utilized to construct technical documents such as design specifications. Brown et al. (Brown et al., 2020), for example, proved the usefulness of GPT-3 in creating cohesive and fluent writing.

GPT models have considerable influence on the design specification process, with a potential to minimize the time and resources needed to develop design specification documents. Engineers and architects can utilize GPT models to enter the essential information required to generate the design specification, and the AI model will produce the remaining portion of the document (Hayman, 2022; Parm AG, 2023). The usage of GPT models can also increase the design specification document's accuracy and consistency. ChatGPT can be used as an interface to prompt designers and engineers for the necessary feedback. The use of GPT models in the design specification process opens up new avenues for automating the development of technical documents. Also, the usage of GPT models can enhance the existing state of manual document generation.

## 5.3 Construction Phase

This section presents the opportunities of leveraging GPT models during the construction phase as shown in Figure 5.

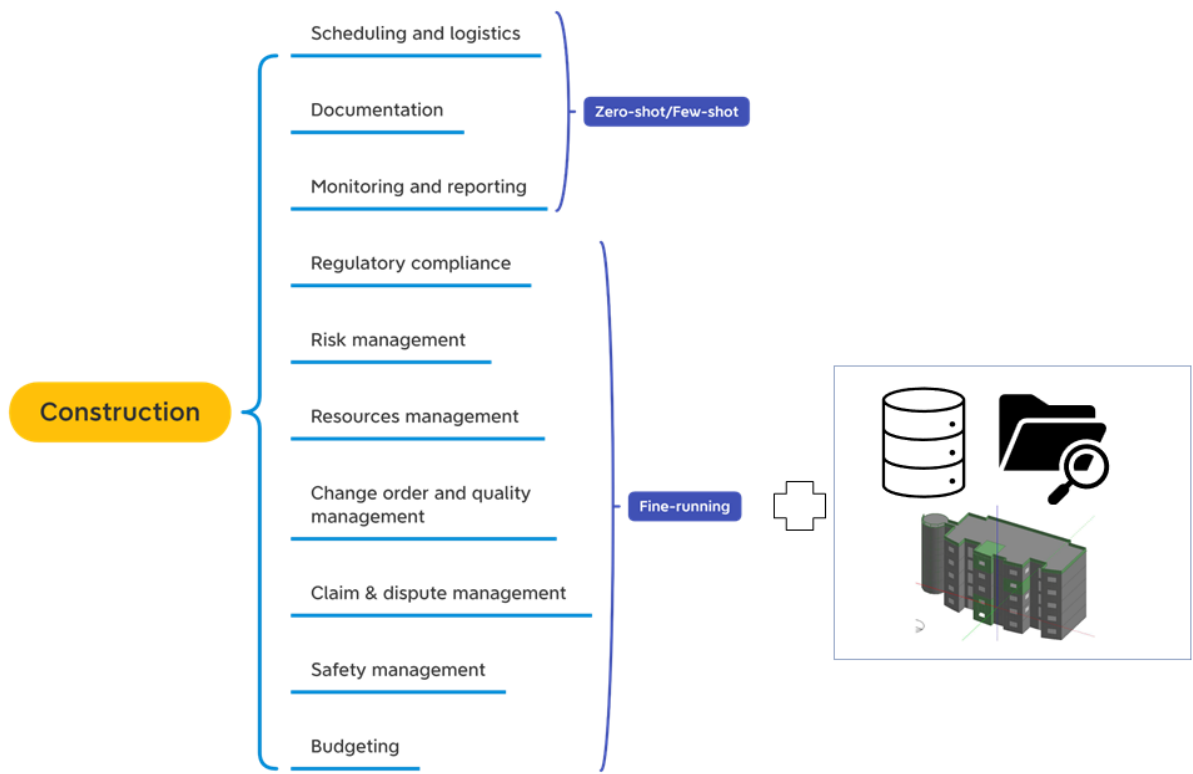

Figure 5: Opportunities for GPT models in Construction Phase

### 5.3.1 Scheduling and logistics management

Scheduling entails developing a comprehensive plan that outlines the sequence of activities, tasks, required resources, and estimated time for completion (Castro-Iacouture et al., 2009). On the other hand, logistics management involves the strategic planning, execution, and control of the movement of personnel, materials, and equipment to and from the construction site (Dannoun, 2022). Efficient scheduling and logistics management are critical to the success of a construction project. However, achieving optimal scheduling and logistics management can be challenging due to the complexity of the supply chain and inherent uncertainties.

Previous studies have attempted to mitigate these challenges by adopting mixed integer programming (MILP) and ML models such as long short-term memory (LSTM) (Al-shihabi & Mladenović, 2022). However, GPT models offer several advantages over these techniques. Unlike MILP and LSTM models, which primarily work with structured and numerical data, GPT models can effectively capture and interpret textual information, enabling a more comprehensive understanding of construction project scheduling and logistics challenges. In contrast to MILP models, which rely heavily on formulating mathematical optimization problems, GPT models operate based on learned patterns and representations from the data. This reduces the need for explicit mathematical modelling and allows for a more flexible and intuitive approach to solving scheduling and logistics challenges in construction projects.

### 5.3.2 Automated Regulatory Compliance

One of the key challenges in the construction industry is the complexity of regulatory compliance. Ensuring compliance with numerous statutory requirements and performance-based regulations is crucial for construction companies (Zhang & El-Gohary, 2016). Traditional manual methods of regulatory compliance checking can be time-consuming, error-prone, and resource-intensive. However, the advent of AI, specifically GPT models, offers opportunities to streamline and automate the regulatory compliance process.

GPT models can be leveraged to automate regulatory compliance in the construction industry through their NLP and understanding capabilities. These models have the ability to analyze and extract information from construction regulatory textual documents, transforming them into logic clauses that can be used for automated reasoning (Beach et al., 2020). By automating the compliance checking process, GPT models can significantly reduce the time and effort required for regulatory compliance assessment and serve as an assistant in automated regulatory compliance.

### 5.3.3 Risk identification, assessment, and management

Identifying and addressing risks proactively can help mitigate potential negative impacts on project objectives. These risks can include external factors, internal issues, technical challenges, or unforeseeable events. During construction, risks can arise from various sources, such as project complexity, resource availability, safety hazards, design changes, and contractual issues (Siraj et al., 2019). The customary procedures of risk management in the construction sector are essential, yet they do have their limitations. These methods heavily rely on the expertise of professionals, which can lead to potential misidentification or improper management of risks if the expertise is flawed or lacking. Furthermore, the subjective nature of risk assessment can create inconsistencies. The process can be resource-intensive and time-consuming, and it may not account for subtle or unprecedented risks due to its static nature and the intricacy of construction projects. Additionally, traditional methods may not fully utilize historical data or predictive analytics, leading to less accurate risk profiles (Baccarini, 2001; Liu, 2013). The nonexistence of standardized systems, documentation, and communication among stakeholders can be problematic, resulting in possible oversights and misunderstandings.

In response to these limitations, GPTs offer significant potential for improving risk identification, assessment, and management in the construction industry (Mills, 2001; OpenAI, 2023). GPT models can be leveraged for comprehensive and dynamic risk identification and assessment by scrutinizing large volumes of project documents and historical data, while also adapting to new information as the project progresses. This ability enables current comprehension of the project's danger profile (Miller, 2016; OpenAI, 2023a; Zheng & Fischer, 2023). Predictive analytics can be leveraged to anticipate potential risks and their impacts, improving the overall accuracy of risk profiles (Cornwell et al., 2022) Finally, GPTs can automate the generation of risk reports and other communications, which can improve clarity and consistency in risk communication and mitigate potential misunderstandings and oversights.

### 5.3.4 Progress monitoring/Project report
During the construction phase, project monitoring is essential for ensuring the successful execution of a project. Effective monitoring allows for timely interventions and adjustments, leading to improved outcomes. On the other hand, project reports serve as crucial tools for communicating project status, achievements, challenges, and other pertinent information to stakeholders (Ibrahim et al., 2009). However, conventional methods of progress monitoring and project reporting have their limitations. These methods often rely on human judgment, which can introduce subjectivity and errors into the process. Moreover, they can be time-consuming and may not provide real-time insights into the project's progress, hindering proactive decision-making (El-Omari & Moselhi, 2009).

Previous research has recognized the significance of ML and CV models in automating progress reporting in order to tackle the challenges mentioned (Ibrahim et al., 2009; Kim et al., 2013). These models leverage advanced algorithms and image processing techniques to analyze construction data, such as images, videos, and sensor data, and extract meaningful insights. However, GPT models, with powerful language models, offer an advantage in this context by combining ML, CV, and NLP techniques. With its ability to understand and generate human-like text, GPT models can enhance progress monitoring and reporting automation by analyzing textual project data, such as project updates, documents, and reports. Hence, GPT models can provide a comprehensive solution for automating progress monitoring and reporting in the construction phase.

### 5.3.5 Site safety management
The construction industry is associated with high rates of accidents and fatalities (Dolphin et al., 2021). Managing site safety is, therefore, vital during the construction phase to prevent accidents, reduce injuries, and ensure compliance with safety regulations. However, traditional approaches to site safety management can be time-consuming and labour-intensive (Tixier et al., 2016).

The integration of GPT in site safety management presents new opportunities for improving safety practices during the construction phase. By leveraging the capabilities of GPT, construction companies can efficiently perform safety audits, automated risk assessments, and get real-time insights into potential hazards (Porter, 2021).
GPT can also assist in identifying high-risk areas on construction sites by analyzing text-based data like project documents, safety reports, and worker feedback to identify potential areas of concern that may not be readily apparent. As a result, propose mitigation measures based on these historical data, documented industry best practices, and regulatory guidelines (Ezelogs, 2023; Togal.AI, 2023). Moreover, GPT can play a crucial role in knowledge sharing and training. Construction workers can access GPT-powered platforms for interactive safety training, where they can receive personalized instructions and guidance on safe practices. GPT's NLP capabilities enable effective communication and prompt responses to worker queries, enhancing the overall safety culture on construction sites. Additionally, GPT can facilitate incident reporting and analysis by automatically categorizing incidents, identifying root causes, and recommending preventive measures, thereby enabling proactive safety management.

The advent of GPT in site safety management holds great potential for the construction industry. By leveraging the power of advanced machine learning and natural language processing, GPT can transform the way safety practices are implemented and managed during the construction phase. The integration of GPT can enhance risk assessment, provide real-time insights, improve incident reporting and analysis, and enable interactive safety training. Construction companies embracing GPT in their site safety management practices will be at the forefront of ensuring safer construction sites and protecting the well-being of their workers.

### 5.3.6 Resource allocation and optimization

Resources on construction projects include plant, personnel, and items necessary to complete tasks. Resources allocation is important and can affect the delivery of projects in terms of cost, time and quality. The allocation of resources is dependent on the diverse features – nature of the project and other key attributes (Kusimo *et al.*, 2019). Approaches such as genetic algorithm and particle swarm optimization have been applied in optimizing resource allocation and levelling. GPT models can also be used in resource allocation and optimization by providing detailed information about the project. The models can provide recommendations by taking cognisant of the project requirements, available resources, project duration, and critical path potential challenges. Also, scenario analysis can be conducted with the GPT model and documentation of the resources' allocation.

### 5.3.7 Change order and Quality management

GPT models can assist the project team when there are changes in the project scope by providing aspects of the project such as cost, time and quality that could be impacted (impact analysis) (Abioye et al., 2021). The models can be employed in reviewing change orders to ensure they are consistent with the contract documentation and for highlighting discrepancies in the change order. Also, the models can be leveraged in the documentation of change orders request, communication and providing recommendations in the event of negotiation and dispute resolution.

Quality control and assurance ensure that the quality of project is in tandem with objectives and established standards whilst also mitigating errors and defects. GPT models can be employed in planning and organizing inspection and testing activities based on defined project requirements and familiarization with principles (quality control and assurance in the construction industry). GPT can be used to identify anomalies in quality data, suggest best practices standards for preventive and corrective actions and assist in quality documentation (management, technical and general procedure manual and policy manual).

### 5.3.8 Construction project documentation

Managing project documentation requires significant effort to ensure that all the relevant documents are up-to-date, accurate, and accessible to all relevant stakeholders (El-Omari &

Moselhi, 2009). This is often done traditionally, which is largely dependent on human judgments, thereby resulting in the infrequent provision of accurate and timely construction data (Ibrahim et al., 2009). Hence, the experts have highlighted the adoption of GPT to mitigate this issue, as the models can be trained to recognize and analyse different types of documents or recordings, extract relevant information from them, and create a centralized database for easy access by all stakeholders. As such, progress reports or status updates can be easily generated based on the extracted data from the project documentation.

### 5.3.9 Claim and dispute resolution

The financial health and integrity of construction stakeholders can be affected by the occurrence of claim and dispute resolution; hence, it is a critical aspect in the construction industry that needs optimum attention. Disputes often arise due to various reasons, such as design errors, construction delays, or cost overruns, and can result in lengthy and costly legal battles (Dikbas et al., 2010).

By leveraging historical data, including change orders and project schedules, GPT models have the potential to make predictions regarding the likelihood of disputes or claims arising in ongoing or future projects. Moreover, the models can generate alternative dispute resolution strategies based on the specific circumstances of a dispute. This is achievable by training them on a diverse range of dispute resolution techniques, including mediation, arbitration, and negotiation (Chaphalkar et al., 2015), thereby enabling them to analyse the details of a dispute and suggest appropriate resolution pathways.

### 5.3.10 Project budgeting and cost planning

It is no doubt that effective project budgeting and cost planning is one of the primary concerns during the construction phase of a project. The construction industry is notorious for cost overruns and budget deviations, which have demonstrated significant financial implications and led to project delays (Wang et al., 2012). Conventionally, project budgeting and cost planning rely heavily on expert knowledge and experience, which can be subjective and prone to errors (Cheng et al., 2009).

However, the emergence of GPT can play a crucial role in optimizing project budgeting and cost-planning processes. By leveraging their machine learning (ML) capabilities, these models can analyse large volumes of project data, including historical cost data, project specifications, and industry benchmarks, to provide cost estimates and forecasts. Although, the ability of the models depends on the prompt quality and the data employed in training and fine-tunning the models.

## 5.4 Operation and Maintenance Phase

This section presents opportunities for leveraging GPT models in the operation and maintenance phase as shown in Figure 6.

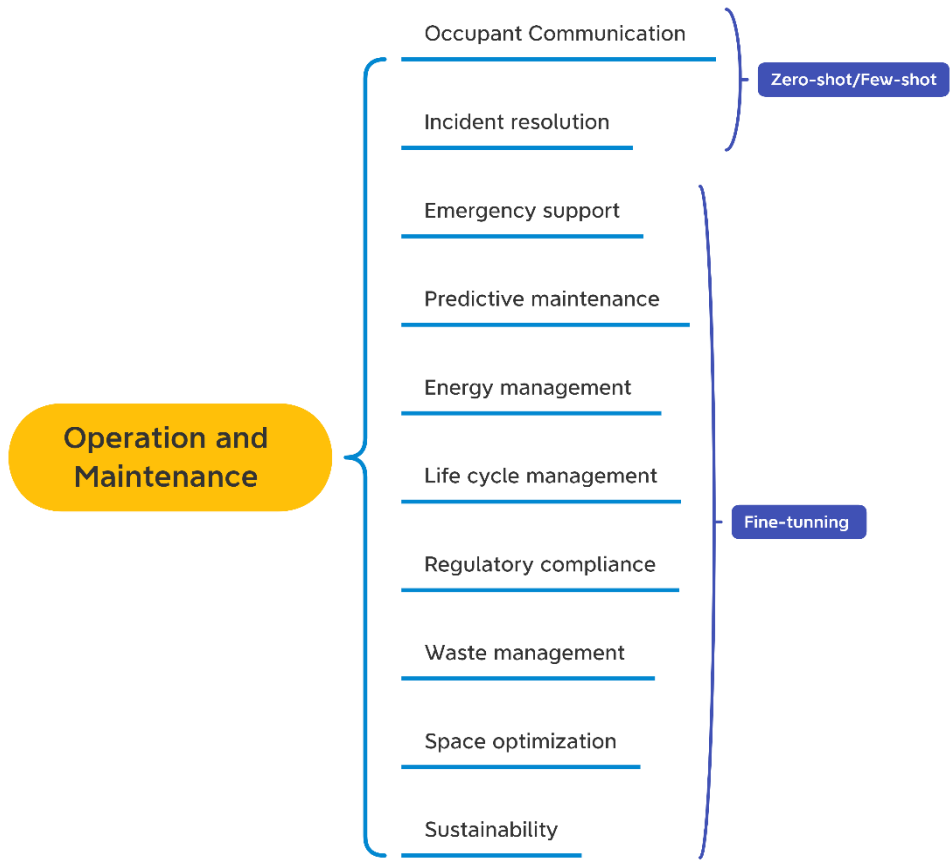

Figure 6: Opportunities for GPT models in Operation and Maintenance Phase

### 5.4.1 Predictive maintenance

One of the key challenges faced during the operational phase of a construction project is the effective management of maintenance activities. Traditionally, maintenance practices have been reactive, relying on scheduled inspections or responding to equipment failures. This approach often leads to unplanned downtime, increased repair costs, and inefficient resource allocation (Bouabdallaoui et al., 2021).

However, with the advent of GPT models, there is a significant opportunity to revolutionize maintenance practices through predictive maintenance. Predictive maintenance involves using data-driven techniques to anticipate equipment failures or performance issues before they occur (Taiwo et al., 2023). By analyzing historical data, sensor readings, and other relevant factors, GPT models can identify patterns and indicators of potential failures. This enables maintenance teams to proactively address maintenance needs, schedule repairs or replacements at optimal times, and minimize the impact on operations. Effective predictive maintenance will help to reduce maintenance costs by avoiding unnecessary repairs and extending the lifespan of the equipment.

### 5.4.2 Energy management and optimization

Buildings and infrastructure are substantial consumers of energy, and ineffective energy utilization not only results in increased operational expenses but also has adverse environmental consequences. Conventional energy consumption management methods primarily depend on manual monitoring and periodic evaluations, which are laborious, resource-demanding, and susceptible to human mistakes (Hagras et al., 2008).

By leveraging GPT capabilities (Zheng & Fischer, 2023), large volumes of data, including historical energy consumption patterns, weather conditions, occupancy rates, and equipment performance, can be analyzed to identify energy-saving opportunities and optimize energy usage. The use of GPT models in energy management and optimization brings numerous benefits to the construction industry. Firstly, it enables real-time monitoring and analysis of energy consumption, allowing facility managers to identify anomalies, detect energy wastage, and take timely corrective actions. Secondly, GPT models can assist in predicting energy demand patterns, allowing for better planning and allocation of energy resources. Thirdly, these models can provide customized recommendations for energy-saving measures, such as adjusting heating, ventilating, and air conditioning (HVAC) settings, optimizing lighting schedules, or upgrading equipment, based on specific building characteristics and usage patterns.

### 5.4.3 Incident reporting and resolution

During the operation and maintenance phase of construction projects, effective incident reporting and resolution are essential for ensuring the safety, functionality, and performance of built assets. Incidents such as equipment malfunctions, safety hazards, or operational disruptions can occur, requiring prompt attention and appropriate actions to minimize their impact. However, traditional incident reporting and resolution methods often involve manual processes, paper-based documentation, and lengthy communication channels, leading to delays in response times and potential miscommunication (Bach et al., 2013). This can result in prolonged downtime and compromised safety measures.

GPT models offer a potential solution to mitigate this challenge. By leveraging the power of NLP and ML techniques, GPT models can efficiently analyze incident reports, identify patterns, and classify incidents based on severity and urgency. In addition, standardized incident reports, capturing relevant details, and facilitating consistent documentation can be generated. This enables stakeholders to prioritize and allocate resources effectively, leading to quicker response times and more efficient resolution of incidents.

### 5.4.4 Lifecycle management of asset

Efficiently managing the entire lifecycle of an asset is crucial to maximizing its performance, minimizing operational expenses, and enhancing overall functionality. Nonetheless, conventional asset management approaches tend to be intricate, time-consuming, and financially burdensome (Grussing, 2014).

In this regard, GPT models offer a compelling solution by streamlining asset management through automated processes encompassing data collection, analysis, and decision-making. By harnessing the power of GPT models, it becomes possible to accurately predict an asset's remaining useful life, empowering maintenance teams to proactively schedule necessary maintenance and repairs, thereby mitigating the risk of unexpected failures. The proactive approach facilitated by GPT models significantly reduces the need for costly emergency repairs and minimizes downtime, resulting in substantial cost savings.

### 5.4.5 Occupant communication and support

GPT models can be employed as a conversational system between the occupant and the facility management (FM) department. These can be prompted to act as chatbots for communicating and supporting facility users in real-time whilst allowing the facility management personnel to attend to meet their requests faster and more effectively. For instance, GPT models can collect maintenance requests, general enquiries from facility users to ensure clear and consistent communication and necessary support. Figure 7 shows ChatGPT prompted to serve as a Facility Chatbot to collect and sort occupant maintenance requests with zero-shot learning.

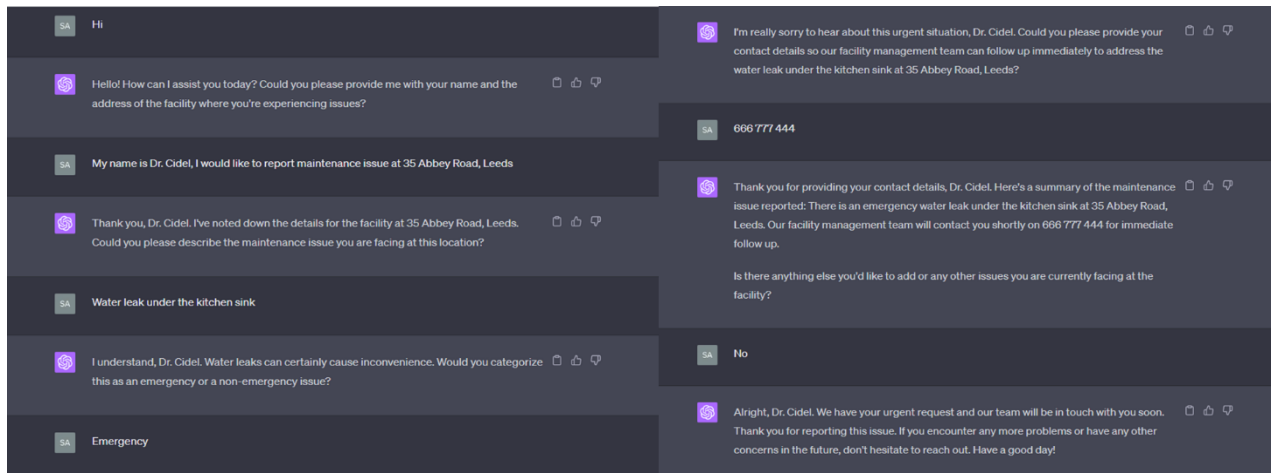

Figure 7: Facility Chatbot

This can be deployed on the FM company's website or an automated voice assistant. For the automated voice assistant, Automatic speech recognition (ASR) and Text to speech synthesiser will be required. ASR will convert speech to text for interaction with the FacilityBot and TTS will convert the text from the system to voice (Saka *et al.*, 2023).

### 5.4.6 Regulatory compliance management

GPT models can be trained with relevant compliance documents, guidelines, and reporting requirements to capture patterns and embedded knowledge. These models can be used in compliance evaluation of facility management activities by comparing new inputs with regulatory requirements to identify non-compliance issues. Similarly, the GPT models can be employed in documentation by highlighting non-compliance issues, providing recommendations, and creating standardized reports to save time and ensure consistency as shown in Figure 8.

Figure 8: Compliance Report Generation

### 5.4.7 Space Optimization and Performance monitoring

GPT models can be leveraged to analyse information about spaces available in a facility to allocate the space for the best usage based on predefined objectives to maximize productivity and utilization. Similarly, data from sensors can provide information about real-time usage to identify peak periods, usage pattern, and congestion areas to optimize space layout, and streamline workflow. This would also enable FM personnel to make space reconfiguration decisions, plan for accommodating future growth, and adapting to seasonal changes and new work trends.

Key performance Indicators (KPIs) such as equipment uptime, user satisfaction, space utilization, energy efficiency, maintenance response time and energy efficiency are some of the metrics employed to measure performance during the facility management phase of construction projects. Information such as maintenance records, equipment performance, energy consumption can be obtained from sensors, project management and energy management systems for usage of GPT models in performance monitoring. Also, real-time data would enable the FM personnel to track performance and deviations from desired performance levels. The developed KPIs can be compared to similar projects and best practices can be recommended by the GPT models. Similarly, performance forecast can be conducted based on historical performance data and GPT models can be leveraged for documentation of facility performance in accordance with defined standard to ensure consistency and saves time.

### 5.4.8 Sustainability reporting and improvement

Sustainability has become a paramount consideration in the construction industry, and stakeholders are increasingly focusing on measuring, reporting, and improving the environmental, social, and economic performance of buildings and infrastructure (Marjaba &

Chidiac, 2016; Taiwo et al., 2023a). However, documenting sustainability practices and metrics accurately presents a formidable challenge due to their intricate and time-consuming nature, requiring extensive endeavours in data collection, analysis, and interpretation.

GPT models offer an opportunity in this area as they can analyze vast amounts of data, including energy consumption records, occupant feedback, maintenance logs, and environmental monitoring data, to provide valuable insights into a facility's sustainability performance. By training GPT models on sustainability frameworks, such as Leadership in Energy and Environmental Design (LEED) or Building Research Establishment Environmental Assessment Method (BREEAM) (Yeung et al., 2020), these models can assist in generating automated sustainability reports. They can extract relevant information from diverse data sources, identify performance gaps, and suggest targeted improvement strategies to enhance the facility's sustainability credentials.

### 5.4.9 Waste management and recycling

Effective waste management and recycling play a pivotal role in the operation and maintenance phase of the construction industry. Properly managing construction and demolition waste, as well as responsibly handling ongoing operational waste, is vital for minimizing environmental impact and fostering sustainable practices (Adedara et al., 2023). Nevertheless, waste management poses a challenge, demanding streamlined processes for efficient tracking, sorting, and disposal methods (Amaral et al., 2020).

Due to their intelligent decision support and optimization capabilities, the experts have identified GPT models to offer valuable contributions to improving waste management and recycling practices. By training these models on waste management regulations, best practices, and recycling guidelines, they can effectively analyse waste-related data and provide informed recommendations. An example of their utility is waste sorting and classification. GPT models can be adapted to analyse images or descriptions of waste materials and based on their training on extensive datasets of waste items and their appropriate recycling or disposal methods, the models can accurately identify the correct handling procedures for various waste types. This capability empowers facility managers to streamline waste management processes and enhance recycling rates, ultimately contributing to more efficient and sustainable practices.

### 5.4.10 Emergency response during fire or other hazards

During fire outbreaks or other hazardous incidents, the ability to respond swiftly and effectively is vital for protecting occupants, minimizing property damage, and restoring normal operations promptly. Emergency situations are often complex and dynamic, demanding rapid decision-making and coordinated efforts among stakeholders (Jiang, 2019).

These complex situations can be addressed using GPT models. This is achievable by training the models on diverse datasets, including information on past fire incidents, emergency protocols, and safety regulations. Subsequently, they can analyse real-time data and provide intelligent decision support to emergency responders. Leveraging their analytical capabilities, these models can assess the severity of the situation, identify potential hazards, and recommend appropriate actions such as evacuation, containment, or fire suppression. Their insights and guidance assist responders in making informed decisions and executing effective emergency plans. The post-incident management can also be enhanced by evaluating incident reports, witness statements, and other relevant documents to identify root causes, lessons learned, and opportunities for improvement (Beata et al., 2018).

## 5.5 Demolition Phase

This section presents the opportunities for leveraging GPT models during construction demolition phase as shown in Figure 9.

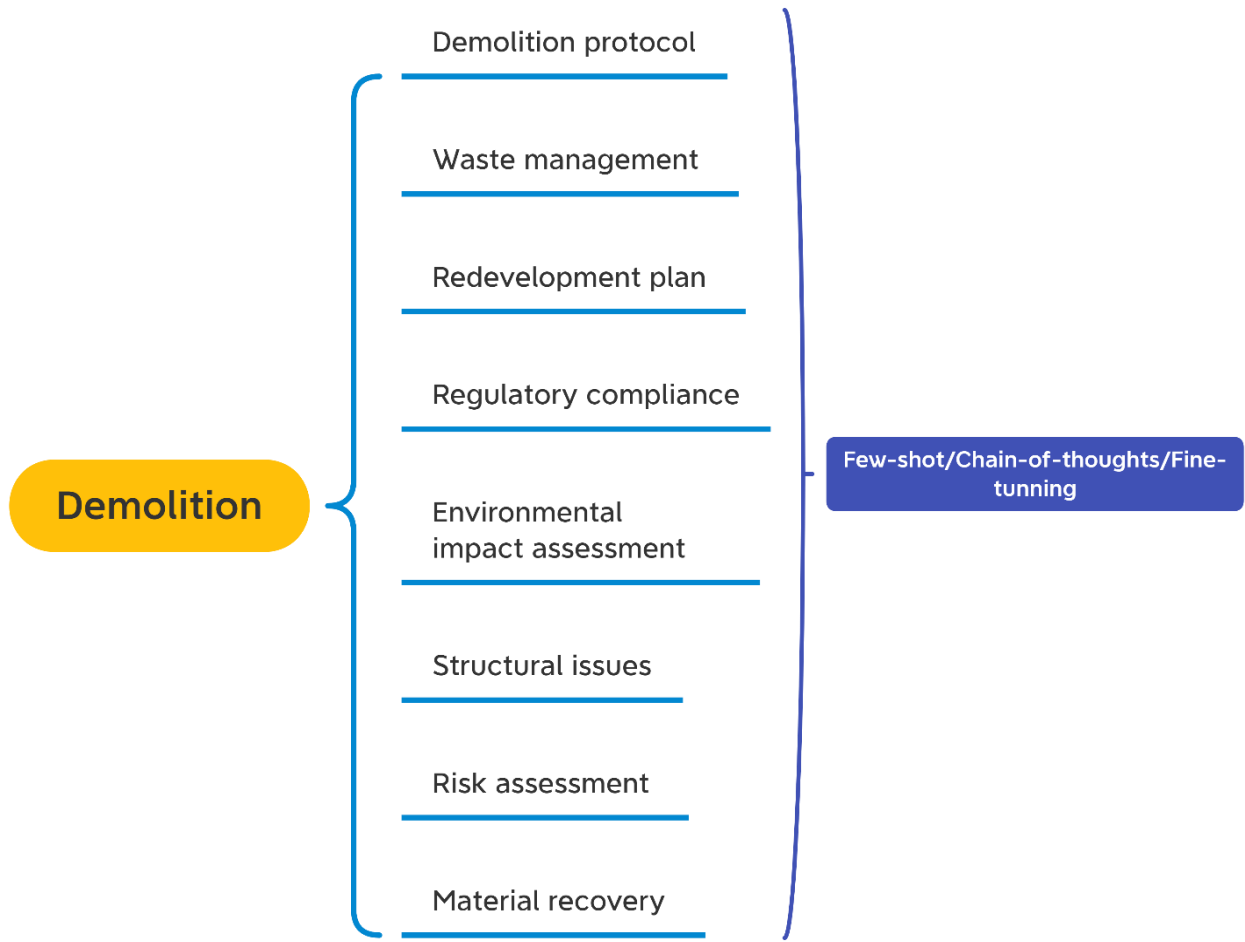

Figure 9: Opportunities for GPT models in the Demolition Phase

### 5.5.1 Development of demolition protocol

The demolition stage of a construction project encompasses the safe and secure elimination of pre-existing structures. It is a multifaceted undertaking that necessitates proper planning and flawless execution in order to safeguard the well-being of workers, preserve the surrounding environment, and achieve successful project completion (Chew, 2010). Developing a robust demolition protocol is crucial to guide demolition activities and mitigate potential risks.

GPT models can play a valuable role in the development of demolition protocols by leveraging their capabilities in data analysis and risk assessment. These models can process large volumes of data, including project plans, site conditions, and historical demolition records, to identify potential hazards, determine optimal demolition sequences, and recommend appropriate safety measures. One application of GPT models in this regard is risk prediction and assessment. Additionally, by considering factors such as site accessibility, waste management strategies, and equipment utilization, these models can generate demolition plans that minimize downtime, reduce environmental impact, and optimize resource allocation.

### 5.5.2 Waste management and recycling, hazardous material management

Demolition projects generate a significant amount of waste, including debris, construction materials, and potentially hazardous substances. Effectively managing this waste and ensuring proper recycling and disposal is crucial for environmental sustainability and regulatory compliance (Rahman et al., 2022). Traditional waste management practices often rely on manual sorting and decision-making processes, which can be time-consuming, inefficient, and prone to human error. The current state of the waste and recycling process can be enhanced by adopting GPT. By leveraging their capabilities in data analysis and pattern recognition, the models can optimize waste sorting, identify recyclable materials, and facilitate the decision-making process for proper disposal of hazardous materials.

### 5.5.3 Redevelopment plan

After the demolition of a structure, a new and promising opportunity unfolds, offering the prospect of redeveloping and repurposing the site. However, developing an effective redevelopment plan can be a challenging task that requires careful consideration of various factors such as site conditions, market demands, and regulatory requirements (Volk et al., 2018).

Conventionally, the development of these plans relies on manual procedures that encompass in-depth research, meticulous data analysis, and extensive consultations with stakeholders. These methods are characterized by their time-consuming nature, high resource requirements, and susceptibility to human biases. Moreover, the intricate nature of integrating numerous variables and constraints often poses challenges when determining the most effective strategies for redevelopment. To address this challenge, GPT models offer valuable assistance in the development of well-informed and efficient redevelopment plans. These models have the capability to analyze vast and diverse datasets, including market trends, demographic information, zoning regulations, and environmental considerations, enabling comprehensive assessments of the site's potential and the identification of viable redevelopment options. Furthermore, GPT models can facilitate scenario modelling, empowering stakeholders to explore different redevelopment strategies and evaluate their potential outcomes.

### 5.5.4 Regulatory compliance and permitting

Adhering to local, regional, and national regulations ensures that demolition activities are carried out safely, minimizing risks to the environment, public health, and worker safety. In the traditional approach, ensuring compliance with regulations and acquiring necessary permits often involves laborious manual procedures, including extensive paperwork, close coordination with regulatory authorities, and meticulous documentation (Macit İlal & Günaydın, 2017). This can lead to delays in the project timeline and increased administrative burden for construction teams. Moreover, the interpretation of regulations and requirements can vary, leading to inconsistencies and potential non-compliance issues.

GPT models have the potential to analyze and interpret regulatory documents, guidelines, and codes to provide real-time guidance and ensure adherence to relevant regulations. They can help identify the specific permits and approvals required for each demolition project, streamlining the permit application process. By automating the analysis of regulatory requirements, GPT models can reduce the potential for errors and ensure a higher level of accuracy in compliance.

### 5.5.5 Environmental impact assessment

The demolition process can have significant environmental implications, including the generation of waste materials, emissions of pollutants, and disturbance to ecosystems. Therefore, it is essential to assess and mitigate these impacts to ensure sustainable and environmentally responsible practices (Uzair et al., 2019).

Typically, conducting environmental impact assessments and implementing effective management strategies have been labour-intensive and time-consuming tasks. They involve extensive data collection, analysis, and compliance with regulatory requirements. The complexity of assessing the environmental impact of a demolition project, considering various factors such as air quality, water pollution, noise, and biodiversity, adds to the challenges faced by professionals in the construction industry.

In this context, GPT models offer promising opportunities to streamline and enhance environmental impact assessment and management processes. These models have the capability to analyze vast amounts of data from diverse sources, including environmental databases, scientific literature, and historical project data. By leveraging machine learning algorithms and natural language processing, GPT models can extract valuable insights, identify potential environmental risks, and recommend appropriate mitigation measures.

### 5.5.6 Establishment of structural issues.

Prior to commencing the demolition process, it is crucial to undertake a comprehensive assessment of the structural condition to identify any potential issues and ensure the safety and efficiency of the demolition operations. The analysis of structural data during this phase can present complexities and consume considerable time. Traditional methods primarily rely on manual inspection, which is subjective, susceptible to human error, and limited in its capacity to handle substantial volumes of data. Additionally, structural problems may not always be discernible through visual examination alone, necessitating a detailed analysis to uncover concealed defects, vulnerabilities, or potential hazards (Xia et al., 2021). These difficulties can result in project delays, escalated expenses, and safety apprehensions throughout the demolition process.

These challenges can be mitigated by the adoption of GPT models. Structural issues such as hidden defects, degradation, or material weaknesses may not be easily detectable through visual inspection alone. GPT models have the potential to identify these hidden defects by analysing historical structural data, maintenance records, and sensor data. By learning from patterns and correlations, these advanced models provide insights into potential areas of concern that may require closer inspection or remedial action.

### 5.5.7 Materials recovery planning and maximisation

Demolition processes often result in significant amounts of waste materials, including concrete, wood, metals, and other valuable resources. Without proper planning and management, these materials can end up in landfills, contributing to environmental degradation and wasting valuable resources (Akanbi et al., 2018). However, identifying and recovering reusable and recyclable materials from the demolition site is usually a complex and laborious task. Conventional approaches rely on manual sorting, which is labour-intensive, error-prone, and limited in its ability to handle large volumes of materials. This hinders the efficient recovery of materials and limits the potential for sustainable resource utilization.

GPT models offer promising opportunities to optimize materials recovery planning and maximize the efficient utilization of resources during this phase. By leveraging their NLP and image recognition capabilities (Zheng & Fischer, 2023), these models can accurately identify materials such as concrete, steel, wood, and plastics. This automation streamlines the process of materials identification, enabling efficient sorting and recovery operations. Another application of GPT in this area is the optimal allocation of the recovered material. The models can be integrated with data on market prices of materials, availability of recycling facilities, and environmental impact assessments. This will enhance the identification of high-value materials for resale, the determination of suitable recycling options, and the proper disposal of hazardous materials.

### 5.5.8 Demolition risk assessment

Processes such as structural dismantling, debris removal, and hazardous material management are associated with demolition activities (Akanbi et al., 2018). Conducting a thorough risk assessment is crucial to identify potential dangers, prioritize safety measures, and comply with regulatory requirements. Traditionally, the risk assessment is performed by relying on expert judgment, which may be subjective and time-consuming (Alipour-Bashary et al., 2022). Furthermore, they may not fully capture the complex interactions and dependencies between different risk factors.

The emergence of GPT offers a valuable solution to enhance demolition risk assessments. GPT models have the potential to analyze various risk factors and provide more accurate and objective assessments by leveraging their advanced NLP and ML capabilities. The training data may encompass demolition project records, structural characteristics, historical accident data, and safety guidelines. This enables the models to identify high-risk areas, predict potential failure modes, and recommend appropriate control measures to mitigate risks.

## 5.6 Value-added Services

These are opportunities that are not directly related to a specific phase of the construction project lifecycle.

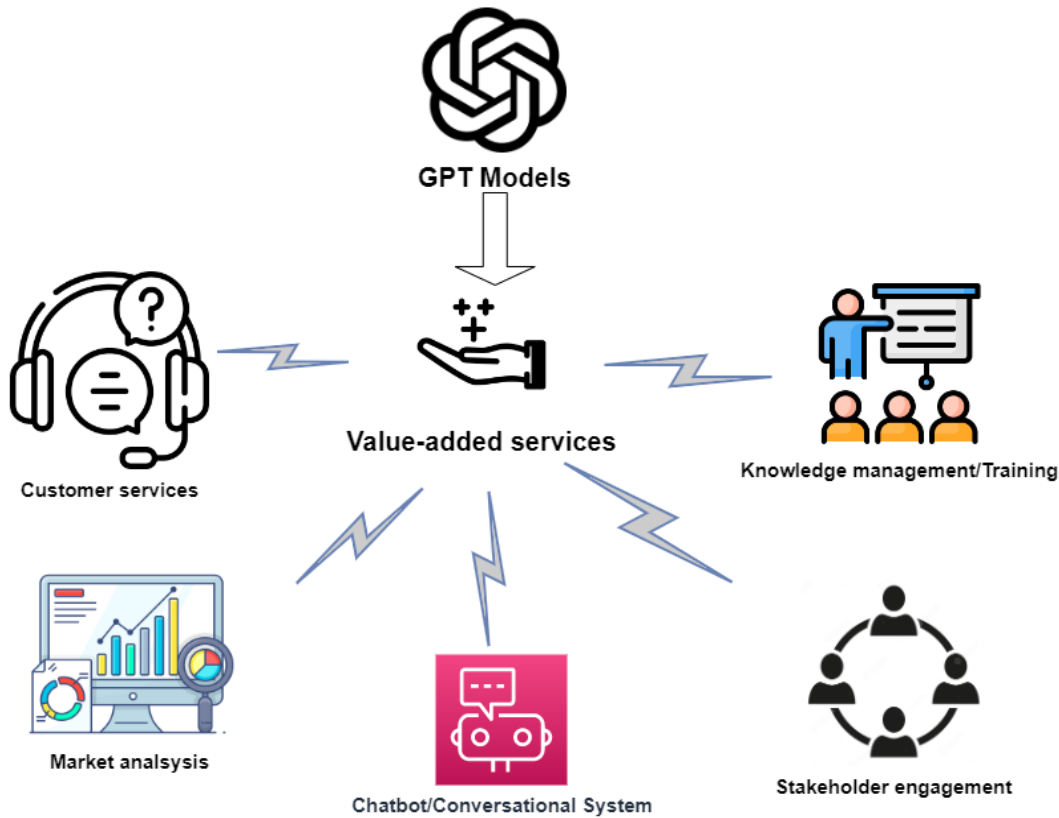

Figure 10: Value-added services

**5.6.1 Knowledge management and training**

Knowledge management and training are perhaps one of the common opportunities for which construction firms can leverage GPT models. Knowledge management deals with identifying, capturing, sharing and dissemination of knowledge and expertise within an organisation (Saka *et al.*, 2022). There is corpus of construction-related information that were used in the training of GPT models, which enable these models to be used for knowledge base creation that could entail best practices, regulatory requirements, safety procedure, design guidelines and project documentation. GPT models enable quick and easy preparation of materials for training the construction personnel on diverse topics such as health and safety and project management. GPT can be deployed as part of the Chatbot system to provide education resources for construction personnel based on their queries, thereby, providing real-time education, personalized training and improving autodidactic experience (Saka *et al.*, 2023).

GPT models can also be leveraged to capture tacit knowledge that is difficult to manage in the construction industry. The experience and expertise of construction professionals can be captured and preserved by using GPT models to gain insights and knowledge from historical data, and past project reports which would enable the creation of knowledge that would benefit new construction professionals. In addition, the construction industry is a global industry, and the construction site often faces language barriers. GPT models can provide multilingual support for knowledge management and training. As such, materials, documents and information can be translated into different languages in a global team and for effective dissemination across language boundaries. This is very important as language barrier often limit productivity and hamper effective health and safety on construction sites.

### 5.6.2 Customer services

GPT models can be leveraged to improve customer experience and satisfaction of firms involved in the delivery of construction projects. GPT models can be used as Chatbot to provide instantaneous assistance to customers and can handle requests and provide information about project progress, product pricing and other general inquires. GPT models can also be employed to provide tailored automated responses to customers' inquiries by identifying the customer's intent from their messages. Similarly, GPT models can enable better customer services across language boundaries and can also be employed in sentiment analysis of feedback from customers. In addition, GPT models can provide personalized recommendations to clients based on their projects and historical data to recommend design options, materials, and relevant construction services. As such GPT models would improve communication, enhance personalize interactions and support that would ultimately contribute to overall customer experience and satisfaction.

### 5.6.3 Stakeholder communication

Diverse stakeholders are involved in the delivery of construction projects, and communication management is the backbone of the effective management of these stakeholders. These stakeholders have different power, interest, and information needs. GPT models can be leveraged for effective stakeholder management in the construction industry to generate project updates to keep stakeholders informed of the project progress, milestones and key developments. GPT models can be integrated with other communication platforms to facilitate personalized interaction and improve stakeholders' engagement. Similarly, GPT models can be prompted for the development of executive summaries of project-related documentation for easy dissemination. Also, GPT models can be employed to analyse feedback from stakeholders and to provide multilingual communication.

### 5.6.4 Market analysis

GPT models can analyse and integrate data related to the construction market, such as industry reports, market reports, new articles, and online posts for trend identification and analysis. In turn, the GPT models can provide insights into changing market dynamics, and demands from customers, emerging technology, and other relevant trends. Also, customer sentiment analysis can be carried out with the GPT models to assist organisations in improving their products and services and aligning with the market. In addition, GPT models can also be employed for market segmentation based on predefined criteria to enable the construction organisation to provide tailored services to the segments. Similarly, they can provide a market analysis of competitors' services, products and marketing, and customer feedback to gain a competitive advantage, and refine their market strategies. Also, based on the historical data market shift, potential market scenarios can be predicted. Lastly, GPT models can help with summarizing market intelligence report to support strategic planning and organisational decision-making, and for monitoring industry news and updates.

### 5.6.5 Conversational System/Chatbot

GPT models can be leveraged in the development of conversational systems/chatbots for natural language recognition, intent classification and entity extraction, and natural language generation. As such, these models would enable conversational interaction via audio or text with the backup application. For instance, GPT models can be employed for information retrieval in BIM via natural language (Zheng *et al.*, 2023) and can also be leveraged in the development of extant systems such as BIM-based AI voice assistant (Elghaish *et al.*, 2022), Voiced-based virtual agents (Linares-Garcia *et al.*, 2022). Leveraging GPT models in conversational system development would improve natural language generation compared to the pattern-based approach currently used in some systems which are rigid and susceptible to errors. Also, GPT models would limit data requirements and engineering expertise necessary for developing prototypes (Zheng *et al.*, 2023)

## 6.0 Challenges

The challenges to the deployment of LLMs in the construction industry could be broadly classified into three categories – industry-related, LLM-related and a nexus of industry and LLM-related as shown in Figure 11.

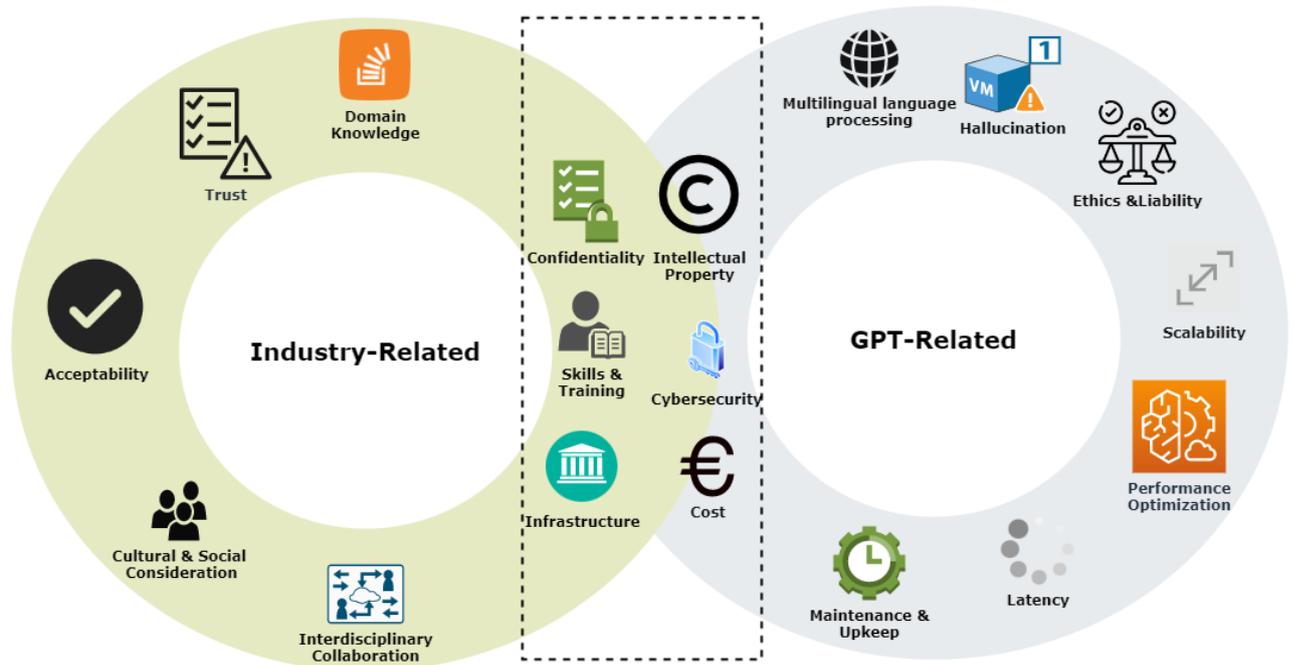

Figure 11: Challenges of GPT models in the Construction Industry

## 6.1 Hallucinations

Despite the improvement in NLP in GPT models, the models are prone to hallucination - give sound and plausible information that are not true- which reduces system performance and users' expectations (Ji *et al.*, 2023). This is highlighted as a major barrier to the application of GPT models in the construction industry by the expert because wrong information could endanger lives and properties, and could also lead to profit loss by companies. For instance, relying on information from GPT models for health and safety management or for programming robotics sequences could lead to accidents on site. Similarly, relying on the hallucinated text on scheduling from GPT models could lead to project delay or cost overrun.

A typical example of hallucinated text from the GPT model is shown in Figure 12.

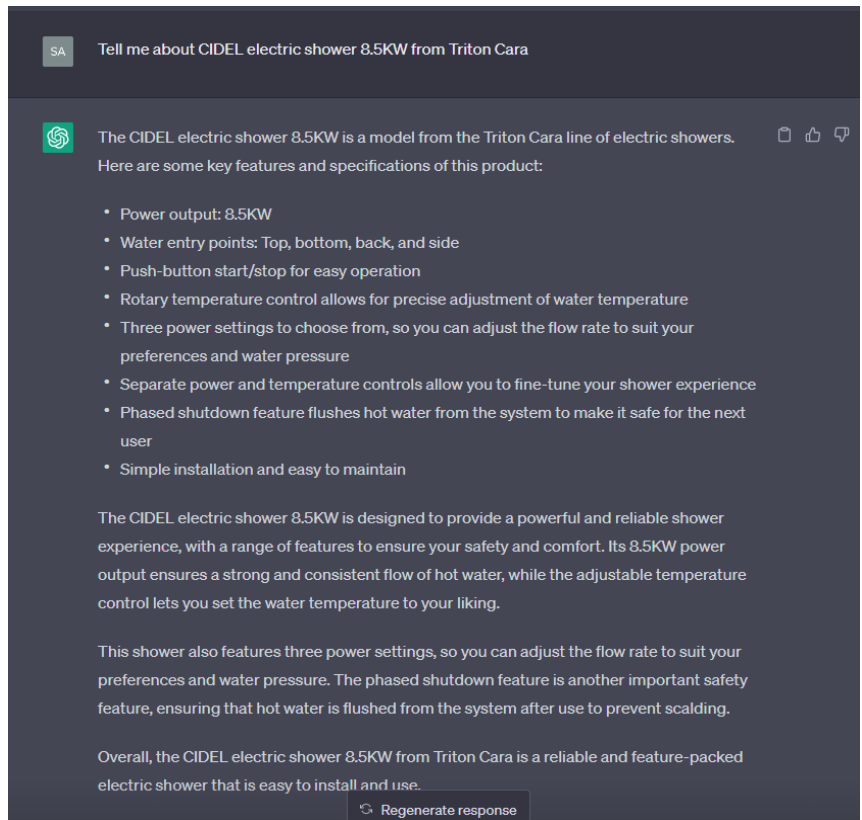

Figure 12: Hallucinated Outputs

This sounds convincing but it is entirely false as there is no CIDEL electric shower from Triton Cara. Rather, CIDEL is a research laboratory 'Construction Informatics and Digital Enterprise Laboratory' at Leeds Beckett University.

### 6.2 Data and Interoperability

Structured data is needed in the fine-tuning of GPT models which are often not readily available in the construction industry. Availability and quality of data have been a major challenge for the application of artificial intelligence in the industry (Abioye *et al.*, 2021). Also, construction projects have unique attributes, and the industry is still slow in the adoption of digital technologies for data collection, leading to potential data loss. This is coupled with the heterogeneity of the data and the time-consuming & labour-intensive exercise in creating labelled data sets.

Furthermore, the cost of interoperability in the construction industry is enormous and ranges to millions of pounds (Shehzad et al., 2021). Datasets are available in different formats such as Portable document format (PDF), doc, hypertext markup language (HTML), Computer-aided design (CAD), Industry foundation classes (IFC) and others. Also, different software and digital tools employed have different formats that makes seamless interoperability difficult. Data are often converted to a standardized form like IFC to improve interoperability but at a loss of watering down the richness of the data (Saka et al., 2023). GPT models currently only accept specific formats (JSON, CSV, JSONL, TSV, or XLSX) which imply that to leverage construction data in GPT models or integrate GPT models with existing tools would involve conversion and pre-processing. For instance, converting BIM model to JSON format, however, not all the information in the BIM model can be converted which poses a challenge to the deployment of GPT models.

### 6.3 Domain-specific knowledge and Regulatory Compliance

Although GPT models are large language models and trained on large data sets, their ability to understand domain-specific knowledge is limited. This is a major barrier in the construction industry which requires a technical understanding of different principles, best practices and regulations. As such, there is a need for adequate fine-tuning of GPT models and provision of context to improve their performance in a technical domain like the construction industry. Similarly, the regulations in the construction industry are many, as it is one of the most regulated industries, due to safety and quality needs. These regulations vary from time to time and based on different contexts such as countries with different environmental regulations, labour, and safety-related laws. Most importantly, these regulations are technically drafted and require technical knowledge and logical reasoning to understand. Consequently, there is currently a need for an adequate breakdown of the regulations for easy understanding of GPT models especially complex pictorial regulations and the need to update the understanding of the models as regulations changes from time to time.

### 6.4 Confidential and Intellectual property

GPT models are trained on large datasets, and these models improve their performance with interaction over time. In the construction industry, the sensitive data such as project design, cost, contracts, and schedules are generated and could be used as inputs for GPT model applications. This raises the concern of confidentiality of GPT models generating output with sensitive information. Also, using data such as project designs and patented building techniques could infringe on the intellectual property owner and leads to ethical and legal issues. Lastly, the ownership of data has always been a problem in the construction industry and the use of GPT models begs the question of who can use data created during the construction project for training large language models? Consequently, there is a need for a clear policy and guidelines for the use of GPT models in construction industry to avoid confidentiality and intellectual property challenges.

### 6.5 Trust and acceptability

The construction industry is known to be slow in the adoption of innovation compared to other industries. There is often resistance to change in the industry, and digital technologies such as BIM have not been widely implemented as expected. With the growing application of AI in the construction industry, industry practitioners and stakeholders are sceptical about trusting ad accepting it. Decisions by stakeholders have significant implications for cost, safety, quality and time of project delivery and relying on GPT models that are 'black box' is difficult. As projects often involve different parties, the issue of acceptance by some parties would hinder the deployment of GPT models on such projects and could also influence the unavailability of project data to fine-tune the models. Also, large language models do require access to large databases for training, with construction firms having concerns about the data usage, and are often unwilling to release such data. This is coupled with the fear that AI would take over jobs in the construction industry and the perceived complexity of digital tools. Consequently, trust and acceptability are inherent challenges in the construction industry that would hinder the applications of GPT models. There is a need to demonstrate the tangibility of GPT models and highlight their reliability through robust testing and validation. It is important to employ an inclusive approach which would involve all stakeholders in considering the usage of GPT models to have clear guidelines and to allay fears.

## 6.6 Liability and Ethics

GPT can be leveraged for many opportunities at the design, construction, management, and demolition stage of construction projects. However, a growing area of concern in the use of AI like GPT models is liability and accountability challenges (McAleenan, 2020). GPT models are LLM trained on a large amount of data which influences what the models would generate during deployment. Bias, incomplete information and inaccuracies in the training data would affect the output generated by the models, and this could cause harm and business loss in the construction industry. Similarly, the inability of GPT models to fully comprehend various legal and regulatory requirements in the industry could lead to the provision of inaccurate and non-compliant advice and recommendations from the system. Also, decision-making process of GPT models is difficult to interpret, and lack transparency and explainability which would lead to a lack of trust and hinder accountability. Stakeholders relying on the outputs or recommendations from the models are unclear about the internal or systematic process employed by the models (You *et al.*, 2023). These beg the question of who is responsible for the potential harm caused by wrong information generated from GPT models. There are currently no clear legal frameworks on accountability and liability for the harm caused by application of AI in construction industry, which could lead to exploitation and leaving affected parties with no clear legal recourse. However, the optimal current approach to overcome liability and accountability challenge is to have clear policies and guidelines for the usage of GPT models, human oversight and to employ models as complement to the human personnel.

Similarly, the problem of ethical issues facing LLM affects GPT models in the construction industry. GPT models have the capability of amplifying biases and discrimination learnt from the training data set, can be used unethically, and could impact the labour market in the construction industry. As such, it is important, for the GPT models to be trained on unbiased and diverse datasets to ensure accurate and reliable results. During the deployment in construction organisations, it is important to have clear guidelines to prevent the model's misuse. In addition, there are economic impacts of GPT models in the industry which could lead to the loss of jobs, pose as a competitor, and could replace repetitive tasks such as information retrieval from texts done by some low-skilled workers in the industry. It is thus important to take steps to evaluate the impact of GPT models in the AEC industry and mitigate negative effects.

## 6.7 Skills and training

Effective deployment of GPT models in the construction industry requires new skillsets and training programs to ensure that the professionals can leverage them properly. The most needed skill is 'prompt engineering' which deals with developing and optimizing prompts to effectively use LLMs (Liu *et al.*, 2023). It is encompassing skills needed to build, interact, and develop new capabilities with LLMs. There are different techniques for prompt engineering such as zero-shot, few-shot and chain-of-thought (CoT) prompting (Wei *et al.*, 2022). In zero-shot prompting, no examples are provided for the models to perform specified tasks whilst in few-shot prompting provide contexts and examples in the prompts for the model to perform better (Wei *et al.*, 2022). On the other hand, CoT prompting provides a 'series of intermediate reasoning steps' which significantly boost the ability of LLMs to perform complex reasoning tasks. Further, there is a need to understand the fine-tuning process of GPT models which enables the models to perform better than prompt design, as the models can learn from more examples than can fit in a prompt and lower latency request (OpenAI, 2023). However, this requires structured training examples in specific formats (JSON, CSV, JSONL, TSV, or XLSX) which implies the need to have the skillsets to pre-process data. As such, skills, and training required for the proper deployment of GPT models in the construction industry serves as bottlenecks.

### 6.8 Infrastructure Requirement and Cost

Leveraging LLMs in the construction industry would require necessary infrastructure such as computing power, network connectivity and data storage. This poses a challenge for the small and medium-sized enterprises which represent about 80% of the organisation in the industry (Saka & Chan, 2020). Also, the current cost of web access for ChatGPT is $20/month with restricted traffic. Similarly, the development of applications that leverage GPT models is not free, as usage is billed per 1000 tokens (~750 words). This ranges from $0.09 / 1K to $0.18 / 1K (prompt and completion cost) for 8K and 32K GPT-4 model (OpenAI, 2023). Gpt-3.5-turbo (ChatGPT) employed for use case validation cost $0.002 / 1K tokens (OpenAI, 2023). Consequently, asides from the infrastructure requirement, the cost of leveraging on LLMs are barriers to deployment in the construction industry. Chen et al (2023) proposed strategies to leverage LLMs at a reduced cost with improved performance. The study highlighted prompt adaption, LLM approximation, and LLM cascade as strategies; and validated 'Frugal GPT' (LLM cascade) with 98% cost reduction.

### 6.9 Scalability and Performance Optimization

Whilst GPT models have overcome some of the scalability challenges inherent in the development of Conversational AI systems (Saka *et al.*, 2023) via a generic model that can be personalized with a prompt design, the challenge of scalability of fine-tuned models persists. Fine-tuning improves learning of GPT models by training on more structured training data for a specific use case. This could be classification (e.g sentiment analysis, email categorization) or conditional generation (e.g entity extraction, chatbot for customer support) (OpenAI, 2023) and would require different datasets. As such, the fine-tuned model would suffer from scalability to support different tasks effectively. Fine-tuned models developed as chatbots for customer support would not be optimal to achieve information retrieval from BIM. Similarly, while the performance of GPT models can be optimized by better prompt designs, other approaches such as self-reflection (Reflexion) require skilled personnel to improve the models (Shinn *et al.*, 2023). Also, fine-tuning models to optimize performance requires structured data and infrastructure, which might be difficult for the majority of the organisations in the construction industry.

### 6.10 Cybersecurity

In an era of digital advancements, cybersecurity has become a critical concern across various industries, including the construction sector (Nyamuchiwa et al., 2022). As the construction industry embraces the potentials of GPT, it also faces significant cybersecurity challenges.

The utilization of GPT within construction processes brings forth fresh areas of vulnerability that can be manipulated by malicious individuals. The dependency on interconnected systems, cloud computing, and data exchange amplifies the likelihood of unauthorized entry, data breaches, and cyber-attacks (You & Feng, 2020). It is imperative for construction firms to confront these obstacles in order to protect sensitive information and uphold the steadfastness of their operations. Alongside external cyber threats, construction companies must also address insider threats. These can arise from unintentional errors or deliberate actions by employees who have access to sensitive information (Nyamuchiwa et al., 2022). Proper employee training and awareness programs are crucial to mitigate the risk of internal data breaches or compromises. It is vital for organizations to foster a cybersecurity-conscious culture and promote best practices among employees. Furthermore, construction companies need to invest in advanced threat detection systems, intrusion detection and prevention systems, and real-time monitoring tools. Prompt response to security incidents is crucial to minimize potential damage and prevent unauthorized access to critical systems or data.

### 6.11 Interdisciplinary collaboration

In the dynamic and complex landscape of the construction industry, successful project outcomes often hinge on effective collaboration among diverse disciplines (Oraee et al., 2019). Interdisciplinary collaboration involves the integration of expertise from various fields, such as architecture, engineering, construction management, and data science, to address the multifaceted challenges encountered throughout the project lifecycle (Fulford & Standing, 2014).

While the introduction of GPT in the construction industry brings promising opportunities, it also presents unique challenges to achieving seamless interdisciplinary collaboration. Professionals from different disciplines often use specialized language and terminology specific to their respective fields. GPT-generated outputs may include technical terms and jargon that might be unfamiliar to professionals from other disciplines. This can lead to misunderstandings, misinterpretations, and inefficiencies in collaborative efforts. For successful interdisciplinary collaboration, trust and acceptance among team members are crucial. The introduction of GPT may raise concerns and scepticism among professionals who are unfamiliar with its capabilities or who fear the potential displacement of their roles. Building trust and fostering acceptance of GPT as a collaborative tool can be a significant challenge, requiring effective change management strategies (Sezgin et al., 2022).

### 6.12 Cultural and social considerations

Construction projects are diverse and often take place in various cultural and social contexts, involving different stakeholders with distinct values, norms, and practices (Zuo et al., 2012). These cultural and social factors can significantly influence the acceptance, adoption, and effectiveness of GPT-driven solutions.

Due to their automation capabilities, the implementation of GPT in the construction industry may raise ethical and social considerations about the impact on employment and job security. The fear of job displacement and the potential loss of skilled labour can create resistance to adopting emerging technologies such as GPT models (Na et al., 2022). Moreover, concerns regarding data privacy, ownership, and security can undermine trust and hinder the widespread acceptance of GPT-empowered interventions. In order to mitigate these challenges, the relevant stakeholders can implement reskilling and upskilling programs to equip the workforce with new skills required to work alongside GPT systems and clearly communicate data handling practices to ensure compliance with relevant regulations.

### 6.13 Latency issue

The latency issue relates to the time lag that occurs between the input of data and the output of results (Yang et al., 2023). In the construction industry, latency can be especially problematic because of the time-sensitive nature of construction projects.

The GPT algorithms require to be fine-tuned on massive amounts of data to achieve their highest level of performance (Zheng & Fischer, 2023). Hence, the training process can be time-consuming. Another cause of latency is the lack of processing power in some systems. GPT algorithms are computationally expensive, requiring high-performance hardware to achieve fast processing times. The cost of such hardware may be prohibitive for smaller construction firms, leading to latency issues. This latency issue can negatively impact real-time monitoring of construction sites, collaborative design, and automated robotic systems, among others.

### 6.14 Maintenance and upkeep

While GPT systems offer numerous benefits, ensuring their continuous operation and optimal performance over time requires diligent maintenance practices and regular updates (Zong & Krishnamachari, 2022). One of the primary challenges in GPT maintenance is system degradation and drift. Over time, GPT models may experience a decline in performance due to changing data patterns, evolving user requirements, or shifting industry standards. This degradation can lead to reduced accuracy, reliability, and relevance of generated outputs, impacting the overall effectiveness of GPT applications in construction processes. To address this challenge, continuous model training and adaptation are crucial.

The integration of GPT into construction processes involves a complex technical ecosystem comprising servers, databases, network infrastructure, and data processing systems. Coordinating and managing these components can be challenging, particularly as the scale and complexity of GPT applications increase. This challenge can be mitigated by putting in place dedicated IT personnel responsible for infrastructure management, regular system health checks, proactive troubleshooting, and prompt response to technical issues.

### 6.15 Multilingual language processing

In an increasingly globalized world, the construction industry is experiencing a growing need for multilingual language processing capabilities. With projects spanning across borders and involving diverse stakeholders, effective communication and understanding of multiple languages have become crucial (Kraft, 2019; Taiwo et al., 2022).

Although GPT models are constantly being updated with more training datasets (Brown et al., 2020), models trained in limited languages may struggle to accurately comprehend and generate content in unfamiliar or less-represented languages. This variability poses a significant hurdle in achieving seamless multilingual communication.

Using zero-shot learning, Lai et al. (2023) conducted extensive experiments on ChatGPT (i.e., powered by GPT 3.5) to investigate its multilingual capacity using various NLP tasks such as summarization, question answering, named entity recognition, and part-of-speech tagging, amongst others. Their result indicated that ChatGPT underperformed in various languages except for English compared to task-specific models built in specific languages. This is due to the fact that a larger percentage of ChatGPT training data was in English (Koubaa et al., 2023). To address the challenge of language variability, it is essential to create and curate large-scale multilingual training datasets that encompass a wide range of languages, dialects, and domains specific to the construction industry. By incorporating diverse linguistic data, GPT models can better understand and generate content in various languages, enhancing their multilingual language processing capabilities.

## 7.0 Case Study Validation: Material Selection and Optimization Platform

To validate the practical application of GPT in the construction industry, we developed a material selection and optimization prototype. This prototype aimed to validate a use case for GPT in the design phase. To facilitate this, a simple BIM model containing 235 building elements was employed as a test case. The development process involves three key steps, each of which serves a specific purpose. Figure 13 shows the system architecture for the use case validation.

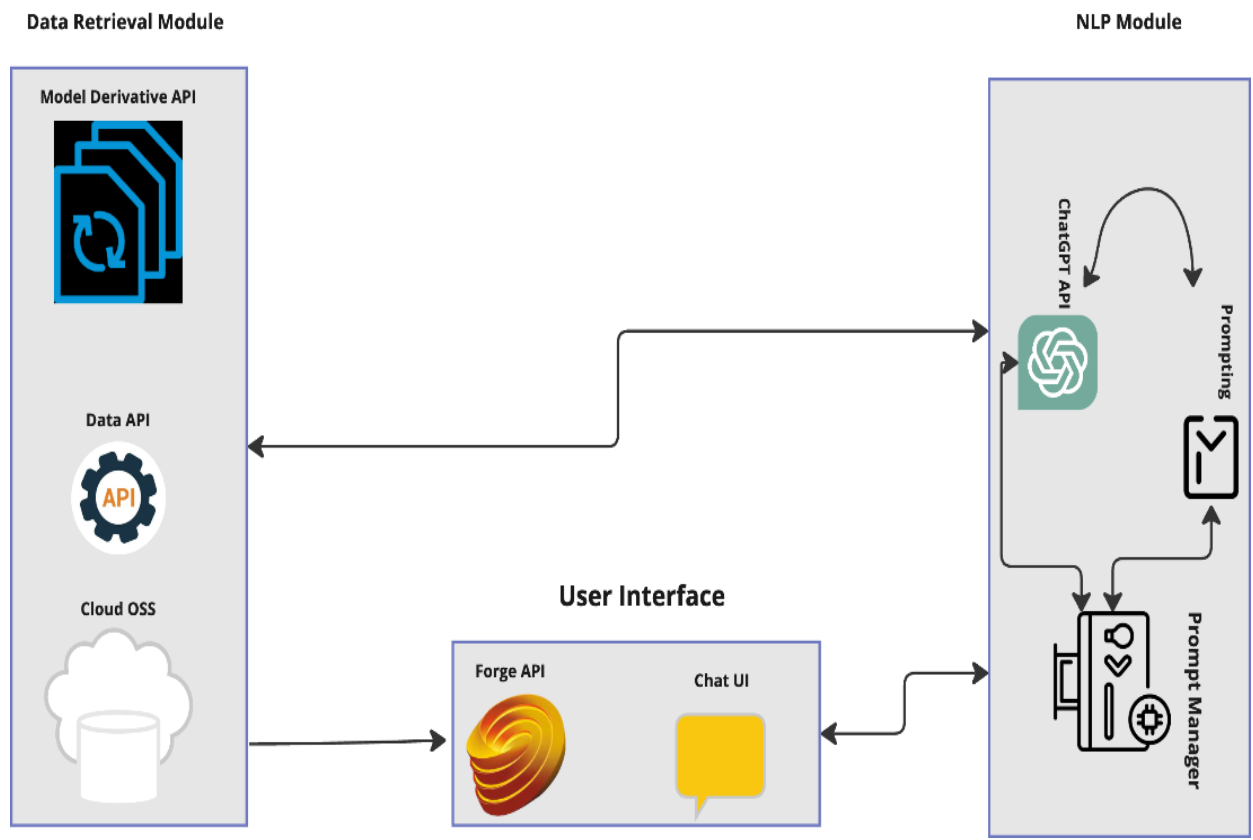

Figure 13: System Architecture

### 7.1 Data retrieval module

The Forge Model Derivative API is used to translate and process the BIM models. The BIM file is uploaded to the forge cloud using the Data Management API and stored for access through the Model Derivative API. The model was then translated into the SVF2 format, enabling the extraction of geometric and metadata information. The Forge Viewer renders 2D and 3D models for web-based access to BIM data. The Model Derivative API extracts data attributes and converts them to a searchable JSON format, including hierarchy trees, geometries, and properties such as component ID, type, and area location. The Data Retrieval module effectively extracts and cleans the necessary BIM data, making it usable for subsequent modules, such as NLP.

```
function onDocumentLoadSuccess(doc) {
  var viewables = doc.getRoot().getDefaultGeometry();
  viewer.loadDocumentNode(doc, viewables).then(i => {
    // documented loaded, any action?
    console.log("loaded");

    getAllLeafComponents(NOP_VIEWER, function (dbIds) {
      AllDbIds = dbIds;
      console.log('Found ' + dbIds.length + ' leaf nodes');
    });

    viewer.addEventListener(Autodesk.Viewing.OBJECT_TREE_CREATED_EVENT, async function (e) {
      //var props = await getBulkProperties(viewer.model, AllDbIds, {propFilter:null,needsExternalId:true,ignoreHidden:false,categoryFilter:null});
      var props = await getBulkProperties(viewer.model, AllDbIds, {propFilter:null,needsExternalId:true,ignoreHidden:false,categoryFilter:null});

      // var searchprops = await searchBulkProperties(viewer.model, AllDbIds, {propFilter:null,needsExternalId:true,ignoreHidden:false,categoryFilter:null});
      var searchprops = await searchProperties(viewer.model, keywords);
      var keywordSearchproperty = await getBulkProperties(viewer.model, searchprops, {propFilter:null,needsExternalId:true,ignoreHidden:false,categoryFilter:null});

      viewer.select(searchprops);
      viewer.fitToView(searchprops);

      //console.log(props);
      // console.log(searchprops);
      console.log(keywordSearchproperty);
      //console.log(AllDbIds);
    });
  });
}
```

Figure 14: Extraction of element property from model derivative API

## 7.2 NLP Prompt Processing Module

The prompt development for material selection and optimization using the ChatGPT API is a multistep process that involves constructing appropriate prompts and designing an interactive dialogue system. The goal is to guide users in selecting optimal materials for each element of the BIM model based on their requirements. The prompt manager connects to the OpenAI API with the ChatGPT model 3.5 turbo using AJAX asynchronous communication. A temperature value of 0.5 is set to reduce the level of randomness in the model's responses.

The prompt consists of two parts: the user prompt and the system prompt. The system prompt is concatenated with the properties of the element in context which is extracted from BIM JSON search before being sent to the ChatGPT API. User prompts serve as the initial input and should clearly express the user's intention, such as requesting assistance in finding the best material for a specific element in the BIM model. Providing sufficient context in prompts helps guide the conversation effectively (Figure 15). System instructions play a crucial role in guiding the behaviour and responses of the ChatGPT. They leveraged the output from the BIM JSON search through the data retrieval module to refine prompts. For example, system instructions may direct ChatGPT to consider factors, such as moisture resistance, when suggesting materials for a bathroom door located at the entrance. These instructions, combined with the user prompts and BIM data search results, ensure that ChatGPT generates relevant and informed recommendations.

```
        // use JSON.stringify to encode myobject to JSON
        const elementpropertyjson = JSON.stringify(elementproperty)

        msgs = [
            {
                "role": "system",
                // You can change this part to be whatever you want it to be
                "content": `Your task is to help a builder select the best optimised material to be used for a selected element in a BIM model using the following step
-Determine if the user input is relevant to material selection for building components
-If No, reply only with "I apologize, but I am unable to handle your request as it is not related to material selection and optimization."
-if Yes then continue with the following steps
1.Element type and location: Determine the type and location of the element from the user input  json `+ elementpropertyjson +`
2.Properties: specify the required properties of the element, such as durability, moisture resistance, and
soundproofing using the location of the element for example doors of toilet require moisture resistance material.
3.Available materials: Evaluate the available materials possible that can be used for the element which includes
materials such as wood, glass, steel, and composite materials.
4.Material comparism: Compare the properties of the available materials to the required properties of the element.
 For example, if moisture resistance is a requirement, materials like PVC, fiberglass, and steel would be better
 options than wood.
5.Other factors: Consider other factors such as cost, sustainability, aesthetics, and energy efficiency when
 selecting a material.
6.Optimal material: Based on the evaluation of available materials and the required properties, choose the
optimal material for the element of the bathroom in the BIM model.
\n`
            },
            {
                "role": "user",
                "content": user_request
            }
        ]
```

Figure 15: Few-shot system prompt instruction

Prompt development is an iterative process that incorporates zero-shot and few-shot scenarios. The performance of the model is continually evaluated, and the prompts, instructions, and dialogue system are refined based on user feedback and real-world interactions. This iterative approach enhanced the efficiency and effectiveness of the system over time.

### 7.3 User interface and integration Module

The user interface (UI) development for the material selection and optimization system involved the use of JavaScript, HTML, CSS3, and Bootstrap 5 (Figure 16). These technologies were utilized to create a chat interface that seamlessly integrates with the forge viewer, which provides a library for visualizing cloud-based BIM. To facilitate smooth communication between the front-end interface and the prompt manager, an AJAX call was implemented. This allowed for the exchange of time-stamped natural language queries and answers as well as the retrieval of result IDs. Using these IDs, cloud-based BIM rendered different 3D contextual scenes of the relevant building objects in response to the user's query, enabling interactive exploration of the uploaded BIM model.

The web-based prototype was designed to be easily deployed to any cloud service provider, ensuring high availability and accessibility from various devices, including mobile devices. The UI seamlessly integrates BIM data available through the OpenAI API, providing a comprehensive and user-friendly experience for material selection and optimization.

```javascript
send_request = function() {

    $("#status_message").html("<span class='sending'>Sending message...</span>")

    // The call to ChatGPT is made from this function:
    CallChatGPT()
};

//Setup forge Viewer
let divId = "MyViewerDiv";
setupViewer(divId, documentId, tokenFetchingUrl, extensionArray);
// Wait for the page to load before we do anything
$("document").ready(function() {

    // Initialize the tabs:
    $("#tabs").tabs();

    // Set initial values from the sliders:
    $("#gpt_temprature_value").html($("#gpt_temprature").val())
    $("#n_gens").html($("#num_gens").val())
    $("#max_tokens").html($("#num_tokens").val())

    $(".min_max").click(function(){

        console.log("Clicked on min/max button")
        console.log("found elements: ", $(this).siblings(".wrapper"))
        // Toggle visibility of the sibling called ".wrapper":
        $(this).parent("h6").siblings(".wrapper").toggle();
    });

    // Capture change of sliders:
    $("#gpt_temprature").on("change", function() {
        $("#gpt_temprature_value").html($("#gpt_temprature").val())
    });

    $("#num_gens").on("change", function() {
        $("#n_gens").html($("#num_gens").val())
    });

    $("#num_tokens").on("change", function() {
        $("#max_tokens").html($("#num_tokens").val())
    });

    // If ctrl + enter is pressed anywhere on the page:
    $(document).keypress(function(e) {
        if ((e.keyCode == 10 || e.keyCode == 13) && (e.ctrlKey || e.metaKey)) {
            send_request();
        }
    });

    $("#send_button").click(function() {
        send_request();
    })

    // Hide the API settings and usage reference:
    $(".api_settings").find(".min_max").click();
    $(".usage_reference").find(".min_max").click();
});
```

Figure 16: Interface set-up and Integration to ChatGPT

To facilitate smooth communication between the front-end interface and the prompt manager, an AJAX call was implemented. This allowed for the exchange of time-stamped natural language queries and answers as well as the retrieval of result IDs. Using these IDs, cloud-based BIM rendered different 3D contextual scenes of the relevant building objects in response to the user's query, enabling interactive exploration of the uploaded BIM model.

The web-based prototype was designed to be easily deployed to any cloud service provider, ensuring high availability and accessibility from various devices, including mobile devices. The UI seamlessly integrates BIM data available through the OpenAI API, providing a comprehensive and user-friendly experience for material selection and optimization.

### 7.4 Discussion

Figure 17 showcases the developed prototype for material selection and optimization. The user interface provides seamless integration of user input queries and ChatGPT responses, both displayed on the right side of the interface. Additionally, the prototype incorporates a visual representation of the 3D model using the Autodesk Forge viewer.

The prototype was tested in 3 different scenarios zero-shot, few-shot with system prompting and edge case prompting scenarios.

### 7.4.1 Zero-Shot prompting

To test the zero-shot capabilities, the prototype disabled the system role prompt in the payload of the ChatGPT API call, retaining only the user prompt and supplying BIM information as input.

Figure 17 show the response generated by chatGPT in response to a user query "Suggest the best material to be used for the door named 'Bois - Panneau de porteto' leading to toilet WC 15" at zero-shot scenario.

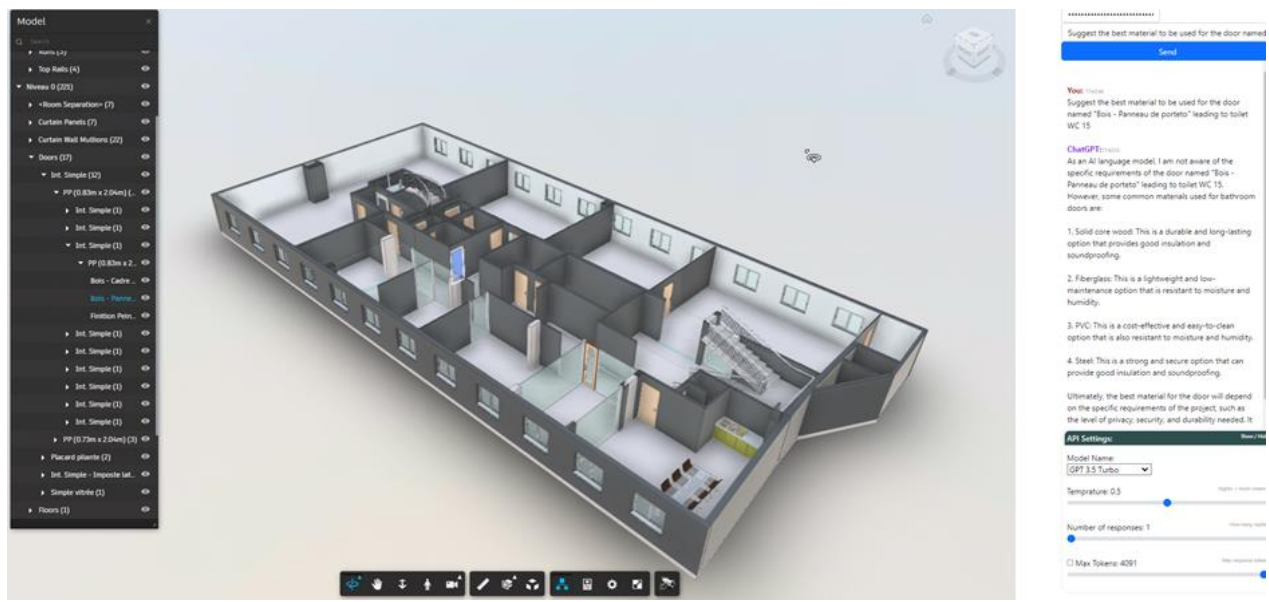

Figure 17: Zero-shot scenario

ChatGPT's response to the query was not specific as it responded with "*As an AI language model, I am not aware of the specific requirements of the door named "Bois - Panneau de porteto" leading to toilet WC 15*" as demonstrated in Figure 8 above. However, it suggested common material for bathroom while laying emphasis on the specific requirements of the user and then recommending consulting professional to determine the best material for the user's specific needs. The above scenario demonstrated that ChatGPT could not serve as an effective tool for material selection and optimisation without proper prompt engineering.

### 7.4.2 Few-Shot Prompting Scenario

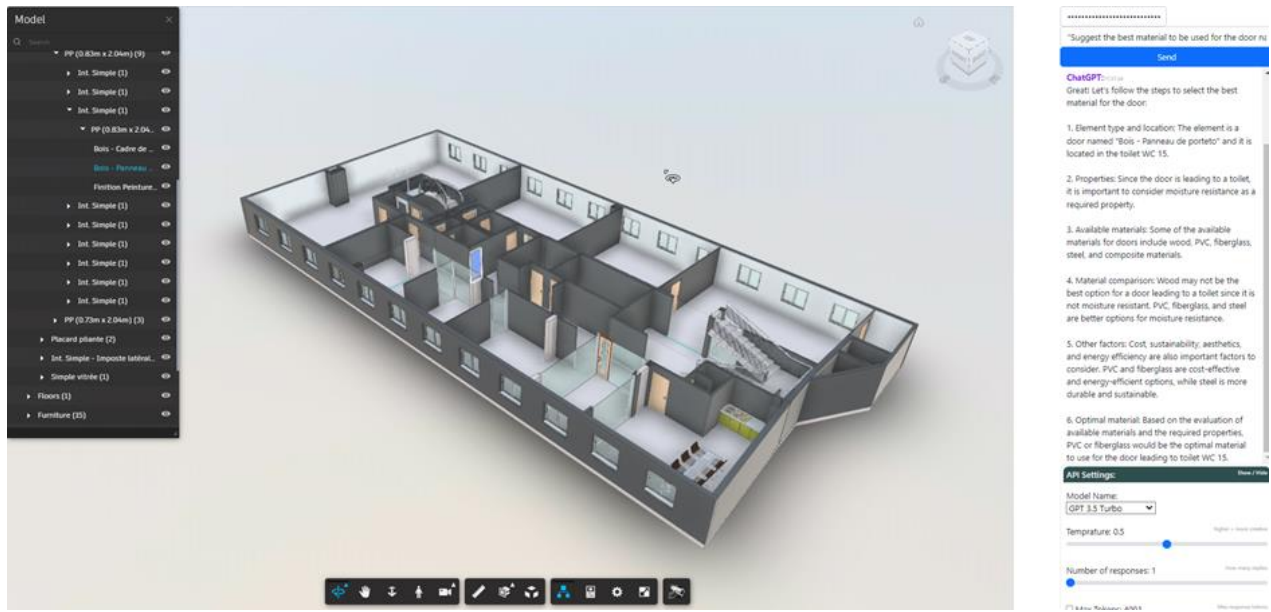

Figure 18: Few-shot scenario

Figure 18 shows the response generated by ChatGPT in a few-shot scenario response to the same user query used in the zero-shot scenario above with the user query: "Suggest the best material to be used for the door to toilet WC 15." ChatGPT successfully provides optimized material recommendations for a bathroom door, taking into consideration various factors. In contrast to zero-shot scenario, it carefully avoids suggesting wood as it is prone to moisture damage in bathroom settings and acknowledges the privacy concerns associated with glass, although glazed glass could provide a privacy solution. Instead, ChatGPT recommends steel and composite materials as viable options, owing to their moisture resistance and durability.

ChatGPT demonstrates its understanding of a specific door element and its location within the BIM model. It recognizes that a door to a bathroom, specifically WC15, is susceptible to water damage owing to its proximity to moisture sources. Based on the available data from the BIM model, ChatGPT analyses possible materials for the door, compares them using factors such as location, area of the component, and other relevant attributes. Furthermore, ChatGPT considers additional factors such as material availability, cost, and energy efficiency during the selection process. The user also has the option to specify the factor that holds the utmost importance, enabling ChatGPT to prioritize that factor in its final recommendation.

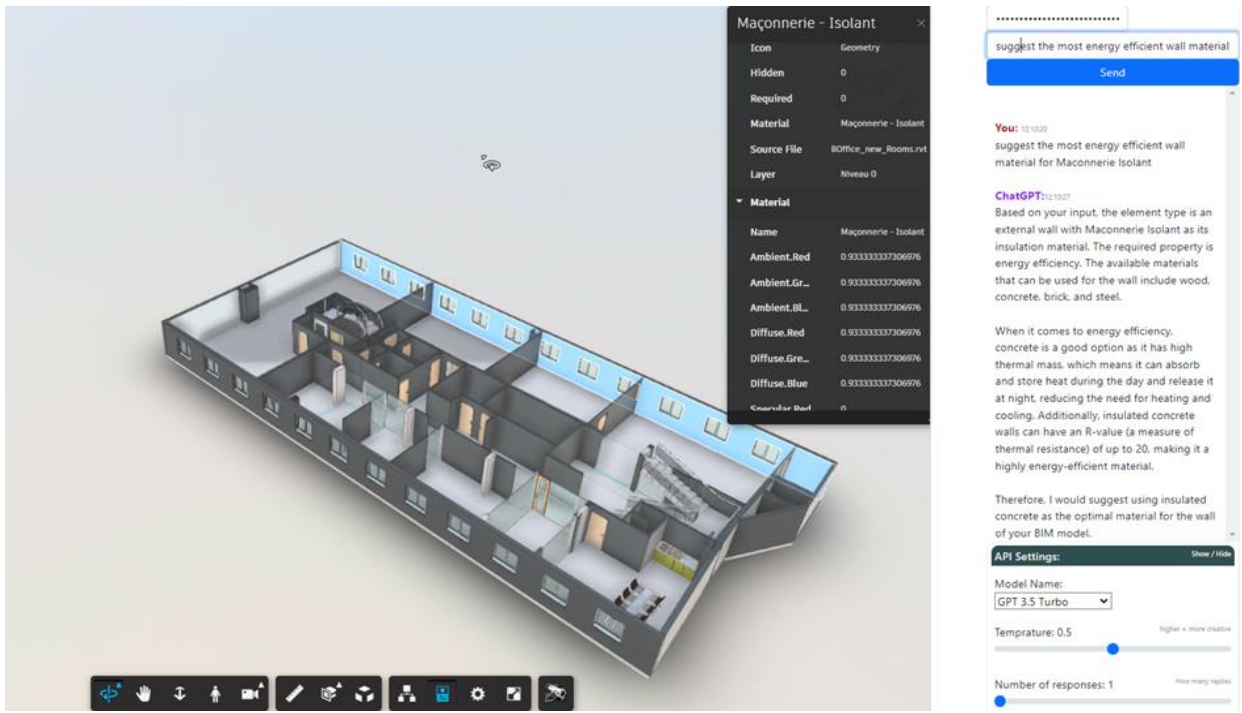

Figure 19: Prompting (Happy path user)

Figure 19 displays another few-shot scenario where the user query is "suggests the energy efficient wall named Maconnerie Isolant." The BIM query, conducted through the data retrieval module, confirms that the wall in question is an external wall before calling up the ChatGPT API (GPT 3.5 Turbo model), which then analyses possible materials, taking into account the available data within the model, such as the external nature of the wall and the user's requirement for energy efficiency to suggest an insulated concrete wall with an R-value of up to 20.

The few-shot scenario showcases the remarkable capability of ChatGPT as a valuable tool for guiding home builders in selecting the most suitable materials for their constructions. By leveraging both general information and user-defined preference parameters, ChatGPT offers informed recommendations that align with the desired characteristics of building components.

### 7.4.3 Edge case Prompting Scenario

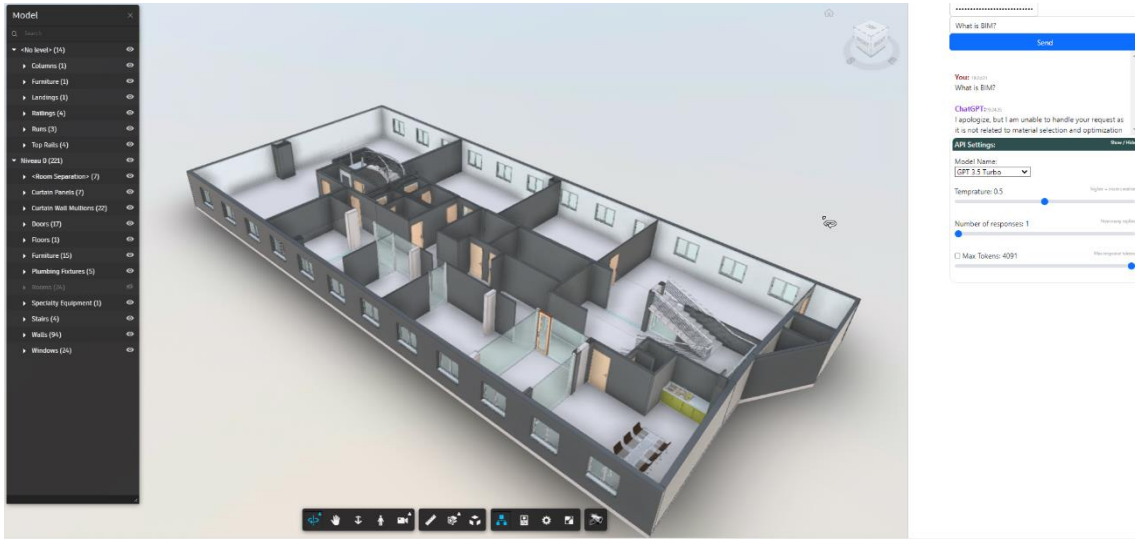

Figure 20: Prompting (Edge case)

In Figure 20, when ChatGPT was prompted with a question '*What is BIM'* that was not related to material selection and optimization, it responded with the message, "I apologize, but I am unable to handle your request as it is not related to material selection and optimization". This indicates that ChatGPT can be tailored to specific needs and deployed as an expert system in a particular field. The prototype allows for easy leveraging of the extensive language capabilities of ChatGPT by readjusting the prompts, without the need for technical experts as required by other NLP systems. This capability empowers construction practitioners to actively participate in rapid testing and prototyping processes, making the system more accessible and user-friendly for industry professionals.

**Limitations**

The prototype demonstrated the capability of selecting the best material for a specific component in a BIM model. However, it requires the identification of the component to provide accurate responses. Future studies would include enhancing the system by probing components and their properties in multiple conversations and leveraging the seamless NLP feature of ChatGPT. This involves guiding the user with a list of parameters to choose from, such as cost, geographical location, and aesthetics, to improve the selection process. Engaging industry stakeholders allows for the collection of query and response data, which can be used to fine-tune the ChatGPT model for better efficiency. In addition, the prompt development process will be evaluated and improved to minimize incorrect or irrelevant responses to user queries. These enhancements will enhance the overall performance and reliability of the material selection and optimization system.

**8.0 Conclusion**

Application of LLMs such as GPT models is still low despite these models overcoming some of the previous challenges that make the deployment of AI difficult in the construction industry. For instance, the time and expertise needed in the development of Conversational AI systems have been drastically reduced by the GPT models trained on large databases. Sectors such as education, medicine, and business have been leveraging these new large language models to improve their modus of Operandi. However, little is known about the opportunities and limitations of leveraging GPT models in the construction industry. Consequently, this study employed three sequential steps to identify and evaluate the opportunities and limitations of GPT models in the construction industry. A detailed preliminary study was conducted and reinforced with an expert discussion on the opportunities and limitations of GPT models. Based on the identified opportunities, a use case was validated, and a prototype was developed for material selection and optimization by integrating BIM and GPT model.

The study reveals that although large language model such as BERT has been gaining attention in the construction industry, GPT models are not well known in the industry until the recent launch of ChatGPT. The study highlights that current applications of GPT models in the industry are for information retrieval, scheduling, and logistics. Opportunities identified for the deployment of GPT models span the predesign, design, construction, and post-construction phases of the project. Also, value-added services are discussed as part of the opportunities for GPT models which include areas that are not directly related to a specific phase of the project lifecycle. These opportunities were discussed, and the findings reveal that the GPT models can be leveraged for some of the opportunities with zero-shot learning, or few-shot learning, or chain-of-thoughts learning via prompt designs. Similarly, some of the opportunities require the need to fine-tune GPT models with structured data to improve the performance of the models and leveraging on existing databases. However, despite these immense opportunities, the study shows that the deployment of GPT models in the construction industry would need to overcome challenges that are inherent in the GPT models and within the construction industry. Inherent limitations of GPT models such as hallucinations, accepted input formats, cost and reliability were highlighted in the study. On the other hand, challenges such as trust, acceptability, domain technicalities, skills, and interoperability which ensue because of the construction industry context were revealed. Furthermore, the study validated a use case with one of the opportunities – materials selection and optimization – by integrating BIM and GPT. The data retrieval module, NLP prompt processing module, user interface and integration module were developed for the prototype. The study shows that leveraging the developed prototype would improve efficiency by providing the stakeholders with materials and optimizing the selection based on defined objectives such as the location of the components and cost.

In addition, there are limitations that serve as fertile grounds for further research. The preliminary search for literature was conducted in Scopus, google scholar, web of science and validated in other databases (ACM and Science direct) which could serve as a limitation. Similarly, the search query developed, and the inclusion criteria of the English language could have led to the missing of some publications. Also, the size of the expert recruited for the discussion could serve as limitation; however, due diligence was observed to ensure the experts have the right experience. The developed prototype was also not subjected to any quantitative validation to evaluate its performance. Further studies can explore the different opportunities areas highlighted in this study by developing prompts, fine-tuning and integrating GPT models with existing databases and repositories. Also, an advanced extension of the materials selection and optimization system is currently being developed by the researchers. This study contributes to the growing body of knowledge on the application of GPT models and provides research vistas for leveraging GPT models in the construction industry. It highlights the current limitations of LLMs and the need to overcome these barriers for the proliferation of LLMs in the construction industry. Its findings would be of benefit to researchers and stakeholders in the AEC industry.

## Acknowledgements

The authors would like to acknowledge and express their sincere gratitude to Leeds Beckett University for providing financial support for this study.